\documentclass[aps,twocolumn,amsmath,amssymb,nofootinbib,showpacs,superscriptaddress]{revtex4-1}
\usepackage[english]{babel}
\usepackage{latexsym}
\usepackage{graphics}
\usepackage{graphicx}
\usepackage{xspace}
\usepackage{epsfig}
\usepackage{color}
\usepackage{bm}
\usepackage{amsmath}
\usepackage{amssymb}
\usepackage{amsthm}
\usepackage{dcolumn}
\usepackage{bm}
\usepackage{float}
\usepackage{hyperref}
\usepackage{color}
\usepackage{epstopdf}
\usepackage{braket}
\usepackage{cleveref}
\usepackage{color,soul}
\usepackage[svgnames]{xcolor}
\hypersetup{hidelinks,colorlinks=true,allcolors=DarkBlue}
\usepackage{multirow}
\usepackage{subfig}

\usepackage{lipsum}  

\usepackage[normalem]{ulem}  

\theoremstyle{remark}

\begin{document}

\preprint{APS/123-QED}

\title{Variational quantum eigensolver techniques \\ for simulating carbon monoxide oxidation}

\author{M. Sapova}
\email{m.sapova@rqc.ru}
\affiliation{Russian Quantum Center, Skolkovo, Moscow 143025, Russia}
\affiliation{National University of Science and Technology ``MISIS'', 119049 Moscow, Russia}

\author{A.K. Fedorov}
\email{akf@rqc.ru}
\affiliation{Russian Quantum Center, Skolkovo, Moscow 143025, Russia}
\affiliation{National University of Science and Technology ``MISIS'', 119049 Moscow, Russia}

\date{\today}
\begin{abstract}
A family of Variational Quantum Eigensolver (VQE) methods is designed to maximize the resources of existing noisy intermediate-scale quantum (NISQ) devices.
However, VQE approaches encounter various difficulties in simulating molecules of industrially-relevant sizes, among which the choice of the ansatz for the molecular wavefunction plays a crucial role. 
In this work, we push forward the capabilities of adaptive variational algorithms (ADAPT-VQE) by demonstrating that the measurement overhead can be significantly reduced via adding multiple operators at each step while keeping the ansatz compact. We reformulate the previously proposed qubit pool completeness criteria to construct complete pools in the tapered qubit space. We also demonstrate numerically that reduction of the qubit pool from polynomial size to linear leads to an increase in the measurement overhead due to the slow convergence of ADAPT-VQE with linear pools.
Within the proposed approach, we simulate a set of molecules, O$_2$, CO, and CO$_2$, participating in the carbon monoxide oxidation processes using the statevector simulator and compare our findings with the results obtained using VQE-UCCSD and classical methods. 
Based on these results, we estimate the energy characteristics of the chemical reaction. 
Our results pave the way towards the use of variational approaches for solving practically relevant chemical problems.
\end{abstract}

\maketitle
 
\section{Introduction}\label{sec:introduction}

Simulating complex quantum systems is among the main potential applications of quantum computing devices~\cite{Lloyd}.
The particular problem is calculating properties of molecules, which is hard to do with a reasonable accuracy (chemical accuracy, defined as 1 kcal/mole, which is required to predict chemical reaction rates) for large-scale systems using existing classical approaches~\cite{McArdle, Bauer_2020, elfving2020quantum,Fedorov2021}.
Quantum chemistry algorithms based on the use of quantum computing consist of two large classes: methods for fault-tolerant quantum computers and noisy intermediate-scale quantum (NISQ) devices. 
Algorithms for NISQ devices take into account {the} limitations of the existing {quantum processors},
so they require low circuit depths that allow executing within the limited coherence time of the device. 
In this context, variational hybrid-classical algorithms~\cite{Cerezo_2021} {are of the particular interest} since they can be implemented even having limited resources of NISQ devices. 
Among existing methods, Variational Quantum Eigensolver (VQE) is a flagship hybrid quantum-classical algorithm for molecular simulations~\cite{Peruzzo}. 
VQE can be seen as a general framework that consists of the following building blocks~\cite{Cerezo_2021}: 
(i) preparation of parameterized state, (ii) cost function estimation, and (iii) parameter optimization. 
While the state preparation and cost function estimation are realized using a quantum device, optimization utilizes classical computers. 
The flexibility of VQE approach lies in the ability to implement both parts of the algorithm, quantum and classical, in diverse ways~\cite{Cerezo_2021, McArdle, fedorov2021vqe}.

As in classical methods of quantum chemistry, the choice of the ansatz for the molecular wavefunction is a crucial step in VQE implementations since it determines the accuracy of the obtained ground state energy.  
The main difficulty in the state preparation is the fact that the expressible and scalable ansatz becomes highly complex and has a long decomposition into quantum gates.
Despite various hardware-efficient ansatzes have been used for simulations of small molecules~\cite{Peruzzo, Kandala2017, Barkoutsos, Gard}, their convergence for larger molecules remains uncertain due to the known problem of barren-plateaus~\cite{McClean}.

For scalable molecular simulations, VQE with chemistry-inspired ansatzes, such as unitary coupled cluster (UCC)~\cite{Romero_2018, Barkoutsos}, is considered as the most reliable since it accounts for the special structure of the molecular wavefunction. 
In contrast to {the} classical projective coupled cluster approach, its unitary variant is more preferable due to the lack of non-variational failure. The restriction of electronic excitations to single (S) and double (D) ones, known as VQE-UCCSD, remains the most popular approach with {many} reported numerical simulations~\cite{Kuhn, armaos2019computational, Rice} and quantum simulations of small molecules~\cite{OMalley_2016, Hempel, McCaskey_2019}. 
In practice, implementations of the UCC ansatz require the use of the Suzuki-Trotter decomposition~\cite{Grimsley_Trot}. Since the circuit depth is one of the major limitations for NISQ processors, the first-order truncation, which neglects the error emerged by non-commuting terms in cluster operators, is a common choice. 
Despite the advantages provided by the UCCSD approach, accounting for single and double excitations is still not expressible enough in the presence of strong correlations~\cite{Cooper_2010, Grimsley_Trot, Lee, Kuhn}. As far as the correlation energy increases with the enhancement of the atomic basis set~\cite{Helgaker_1997, Varandas_2018}, proper accounting for correlation effects is crucial for practically relevant molecular simulations.

A straightforward way to improve the accuracy of computations is to include triple excitations in the ansatz~\cite{Musia2001CoupledCS}.
In classical computational chemistry, such extension leads to significant difficulties, so that is why CCSD(T) --- coupled cluster method with perturbative inclusion of triplets --- is considered as a gold standard providing a trade-off between the computational cost and precision~\cite{Helgaker2001}. 
For quantum processors, even a full implementation of the UCCSD circuits on a quantum computer is currently infeasible due to the sufficiently large gate count. Including triple excitations would lead to a huge overhead in the number of excitations scaling as $\mathcal{O}(N^3n^3)$, where $N$ and $n$ are numbers of spin-orbitals and electrons, respectively. This means that such an extension is not suitable for existing or near-future devices.

A possible solution is to use adaptive methods, such as ADAPT-VQE~\cite{Grimsley}, which iteratively construct the ansatz. At each step of {the} ansatz growing procedure, ADAPT-VQE adds a new operator from a fixed pool.
The ansatz is built according to the gradient criterion, which approximates the idea of accounting for the maximal amount of the correlation energy by adding the operator with the largest gradient. In the reported numerical results, ADAPT-VQE is able to find a pathway to the desired state~\cite{Grimsley, Tang_2021, shkolnikov2021avoiding, Liu2021AnEA, Evangelista_2019}. ADAPT-VQE simulations of the H$_6$ molecule, a prototype of strongly correlated molecules, have been demonstrated by Grimsley et al.~\cite{Grimsley}. 
These results demonstrate that ADAPT-VQE outperforms VQE-UCCSD in terms of the accuracy for the considered example.
The approach appears to be a useful tool for simulating molecules with strong correlation. We assume that calculations of larger molecules are needed to better understand the limitations of the algorithms.

In this work, we consider two strategies for selecting the operator pool: fermionic and qubit ADAPT-VQE approaches. Fermionic ADAPT-VQE uses UCCSD pool, whose size grows as $\mathcal{O}(N^2n^2)$. Qubit ADAPT-VQE instead consists of separate Pauli strings~\cite{Tang_2021}
{making} it closer to the idea of hardware-efficient ansatzes and attractive from the circuit depth point of view. For both approaches, we use ADAPT-VQE procedure with qubit tapering~\cite{Setia_2020} to reduce the complexity of computations. This {procedure} greatly simplifies both the quantum and classical parts of the algorithm.

The major shortcoming of ADAPT-VQE is an overhead in the number of measurements on quantum devices compared to VQE-UCCSD. It consists of two parts: (i) computing derivatives in order to select an operator with the largest gradient from the pool and (ii) the parameter optimization procedure. The number of computed energy derivatives at each step equals the operator pool size. 
Here we propose to add multiple operators at each step of the ansatz growing procedure. We refer to this implementation as ``batched ADAPT-VQE'' and demonstrate its efficiency for both fermionic and qubit ADAPT-VQE implementations. Our numerical investigation demonstrates that such a procedure allows one to significantly reduce the number of computed derivatives in the gradient evaluation part of the algorithm and at the same time leads to insignificant losses in terms of circuit efficiency.

Additionally, we focus on the qubit ADAPT-VQE procedure. In the original paper~\cite{Grimsley}, the qubit pool is built by cutting fermionic excitations into individual Pauli strings, which leads to enlarging operator pool compared to the UCCSD one. However, the numerical results indicated the redundancy of such a pool: a partial removal of the pool operators does not affect the convergence. Recently, Shkolnikov et al.~\cite{shkolnikov2021avoiding} proposed a strategy of building complete pools for qubit ADAPT-VQE~\cite{Tang_2021}, which grow linearly with the system size. In our work, we reformulate the completeness criteria for the molecules in the reduced qubit space after applying the qubit tapering procedure. Moreover, we propose an automated procedure for building linear and polynomial complete pools without manual analysis of molecular symmetries. We compare qubit ADAPT-VQE performance for linear and polynomial pools and discuss the difficulties encountered during the optimization procedure.

For numerical analysis, besides the common test molecules (H$_4$, LiH, and H$_2$O), we simulate a set of industrially-relevant molecules using the proposed method, such as O$_2$, CO, and CO$_2$, and compare the results obtained using VQE-UCCSD, ADAPT-VQE, and classical methods. 
These molecules are involved in carbon monoxide oxidation: CO$+{1}/{2}$O$_2$ $\rightarrow$ CO$_2$.
CO is a common product in partial oxidation of carbon-containing molecules~\cite{Soubaihi_CO_oxid, Dey}. 
As carbon monoxide is highly toxic for humans, the process of eliminating carbon monoxide emissions attracts much attention.
Catalytic oxidation of CO into CO$_2$ as a typical process for pollution reduction is of great industrial interest.
In addition, our choice of the molecules is motivated by their following properties: (i) large correlation energy in the minimal basis set, (ii) set contains both singlet and triplet molecules, and (iii) computations of the considered molecules using simulators of quantum computing devices require less than 20 qubits making it possible to run experiments in a reasonable time. For better understanding the convergence of the considered methods, we estimate the electronic energy of CO oxidation reaction. 
Recently, carbon neutrality has become a global trend, and as part of this, the consideration of e-Fuel and other products is accelerating.
In that sense, chemical reactions via CO$_2$, CO, O$_2$, H$_2$O, etc. are of great importance for the future studies.

\section{Results}\label{sec:statevector_simulation}

\subsection{Batched ADAPT-VQE}\label{subsec:batched_ADAPT}
\begin{figure*}[htbp]
\centering
    \includegraphics[width=160 mm]{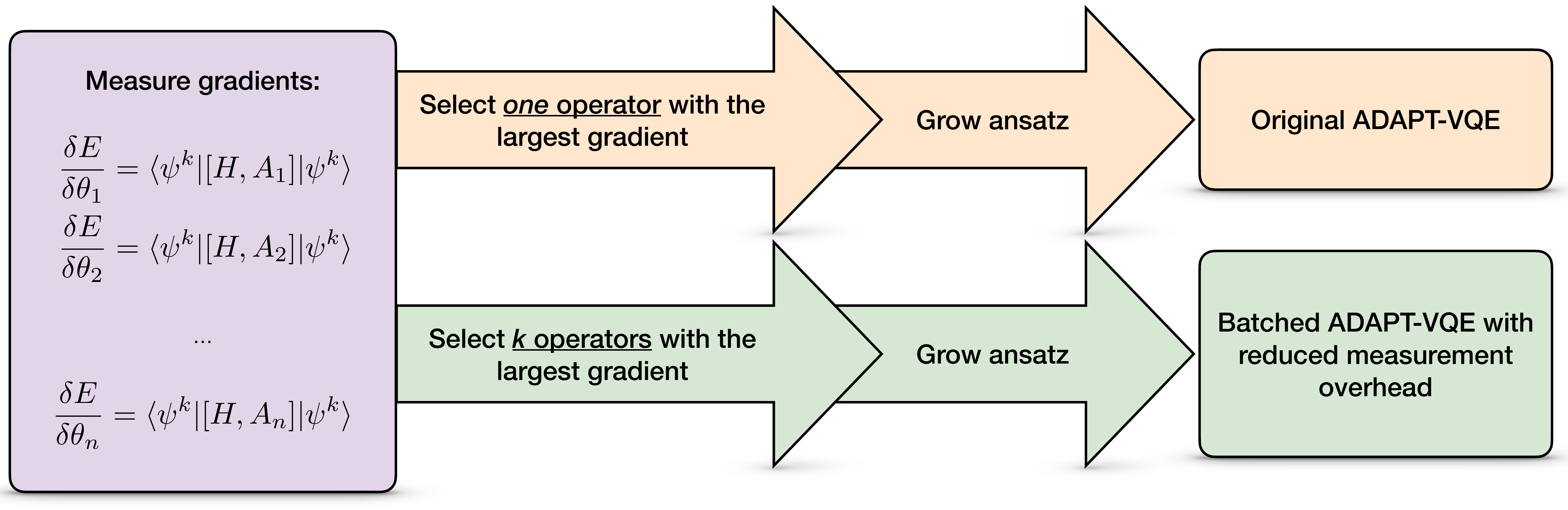}
    \centering
    \caption{\textbf{Scheme of the ansatz growing procedure in original and batched ADAPT-VQE.} Original ADAPT-VQE adds a single operator with the largest gradient value at each iteration. In our implementation of batched ADAPT-VQE, we add multiple operators with the largest gradients at each step of the ansatz growing procedure; $k$ is the number of gradients that differ from the largest one by less than $r$ time}
    \label{fig:example-figure}
\end{figure*}

The ADAPT-VQE is a greedy iterative approach for building an efficient compact ansatz for approximating the molecular wavefunction~\cite{Grimsley}. 
In the original implementation of ADAPT-VQE~\cite{Grimsley}, a single operator with the largest gradient is added at each iteration. 
Such a construction of the ansatz is based on the gradient criterion, which approximates the idea of recovering most correlation energy.
Indeed, the numerical investigation of ADAPT-VQE and randomly constructed ansatzes demonstrates significantly worse convergence of the latter with respect to parameter number in the quantum circuit.
However, computations of the derivatives for the whole operator pool in order to add a single operator produce a notable overhead in the number of measurements. Due to the rather slow convergence of ADAPT-VQE for strongly correlated systems and the increasing size of the operator pool, the application of ADAPT-VQE for practically-relevant molecules is hindered.

In order to reduce the number of gradient computations, 
we introduce batched ADAPT-VQE that adds multiple operators with the largest gradients simultaneously. 
This approach allows reducing the number of gradient computations while building a compact ansatz.

We apply batched ADAPT-VQE with both fermionic and qubit pools. The quantum circuit in fermionic ADAPT-VQE algorithm is obtained from the predefined operator pool, which consists of excitation operators. Note that here we define fermionic operators as in conventional UCCSD instead of originally used spin-adapted operators (see Methods,~\Cref{sec:methods}). As the pool size scales as $\mathcal{O}(N^2n^2)$, fermionic ADAPT-VQE introduces a polynomial overhead in the number of measurements compared to VQE with fixed circuits. Thus, adding multiple gradients to the pool has a practical advantage in both numerical simulations and experiments on a quantum computer.

In addition to fermionic ADAPT-VQE, we also investigate qubit ADAPT-VQE performance. Tang et al.~\cite{Tang_2021} proposed the qubit ADAPT-VQE approach, which builds a qubit pool from individual Pauli strings. Unlike qubit coupled cluster approach~\cite{Ryabinkin_2018, Ryabinkin_2020}, these Pauli strings are obtained by splitting each fermionic operator into subterms with all $Z$-Paulis being removed. This method of qubit pool construction leads to an increase in the pool size compared to the fermionic one. However, it was demonstrated that such qubit pool contains redundant operators and can be reduced to a pool of linear size. In the present work, we study qubit pools of both polynomial and linear sizes. Our numerical investigation demonstrates that even though reduction of the qubit pool size aims to reduce the measurement overhead, in practice polynomial pools require fewer derivative evaluations due to the slow convergence of ADAPT-VQE with a linear pool. Our numerical results illustrates that the use of the polynomial qubit pool with batched ADAPT-VQE is a more efficient strategy compared to implementations of qubit ADAPT-VQE with a linear pool.

Liu et al.~\cite{Liu2021AnEA} added {a} fixed number of operators to improve fermionic ADAPT-VQE slow convergence caused by inaccurate identification of operators due to Valdemoro's reconstruction of 3-RDM. The size of the batch in their implementation should be carefully tuned as inappropriate size leads to significant increases in the circuit depth. 
In our approach, we improve the original ADAPT-VQE approach by adding batches of operators {of the} varying {sizes} at each {algorithm} step.
Note that we add the operators to the ansatz following the order of computed gradients.
The adaptive size of the batch allows adding more operators at the beginning of the ansatz growing procedure when the energy difference is significant, and on the contrary, reduce the batch size when the ansatz is closer to the desired state.
The numerical results demonstrate that such an implementation makes it possible to build an ansatz close to the original one while reducing the computational cost.

At each ADAPT-VQE iteration, we pick all the gradients that differ from the largest by a ratio less than $r$, where $r$ is incorporated as a hyperparameter. We set $r = 2$ for all the considered molecules and our numerical results suggest that batched ADAPT-VQE produces circuits close to the ones obtained with original ADAPT-VQE. From the practical point of view, it is important that the hyperparameter value can be the same for different molecules.
 
Since batched ADAPT-VQE adds multiple operators at each step, it requires sizably fewer iterations to build an ansatz, which considerably reduces the cost of computing gradients with respect to operators in the pool. 
This is also important for classical simulations since computations for CO$_2$ molecule (19 qubits) take a significant amount of time.
We note that in the case of experiments on {existing} quantum devices, 
the restriction of the operator batch size at each iteration can be useful due to the controllability of the quantum circuit size. 
The performance of ADAPT-VQE under noisy conditions requires a separate study.

\subsection{Complete pools}\label{subsec:mcp}

{The initial formulation of qubit ADAPT-VQE considers pools of Pauli strings originated from UCCSD excitations.}
As such strings are obtained by cutting each fermionic excitation and removing $Z$ Pauli operators, {the pool grows} polynomially with the system size as $\mathcal{O}(n^4)$. Theoretically, this produces an $\mathcal{O}(n^8)$ overhead in the number of measurements in the straightforward implementation~\cite{Liu2021AnEA, shkolnikov2021avoiding}. That is why the idea of restricting the pool size seems to be attractive.
We start our discussion with the papers on qubit ADAPT-VQE~\cite{Tang_2021, shkolnikov2021avoiding} that propose the concept of minimal complete pools (MCPs) and prove their existence.

MCPs include a set of operators, individual Pauli strings, that can transform a reference state to any real state in the N-qubit Hilbert space. Tang et al.~\cite{Tang_2021} obtained a 3-qubit pool analytically and proved by induction that minimal complete pools exist for any $N$ and have a size of $2N-2$. 
The authors derived a set of generators forming a theoretically complete pool, which performs successfully on random Hamiltonians. 
However, the derived pool is not appropriate for ADAPT-VQE molecular simulations as the Pauli strings used as generators do not account for specific symmetries presented in {the} molecular Hamiltonian.

\begin{figure*}[htbp]
\centering
    \includegraphics[width=\textwidth]{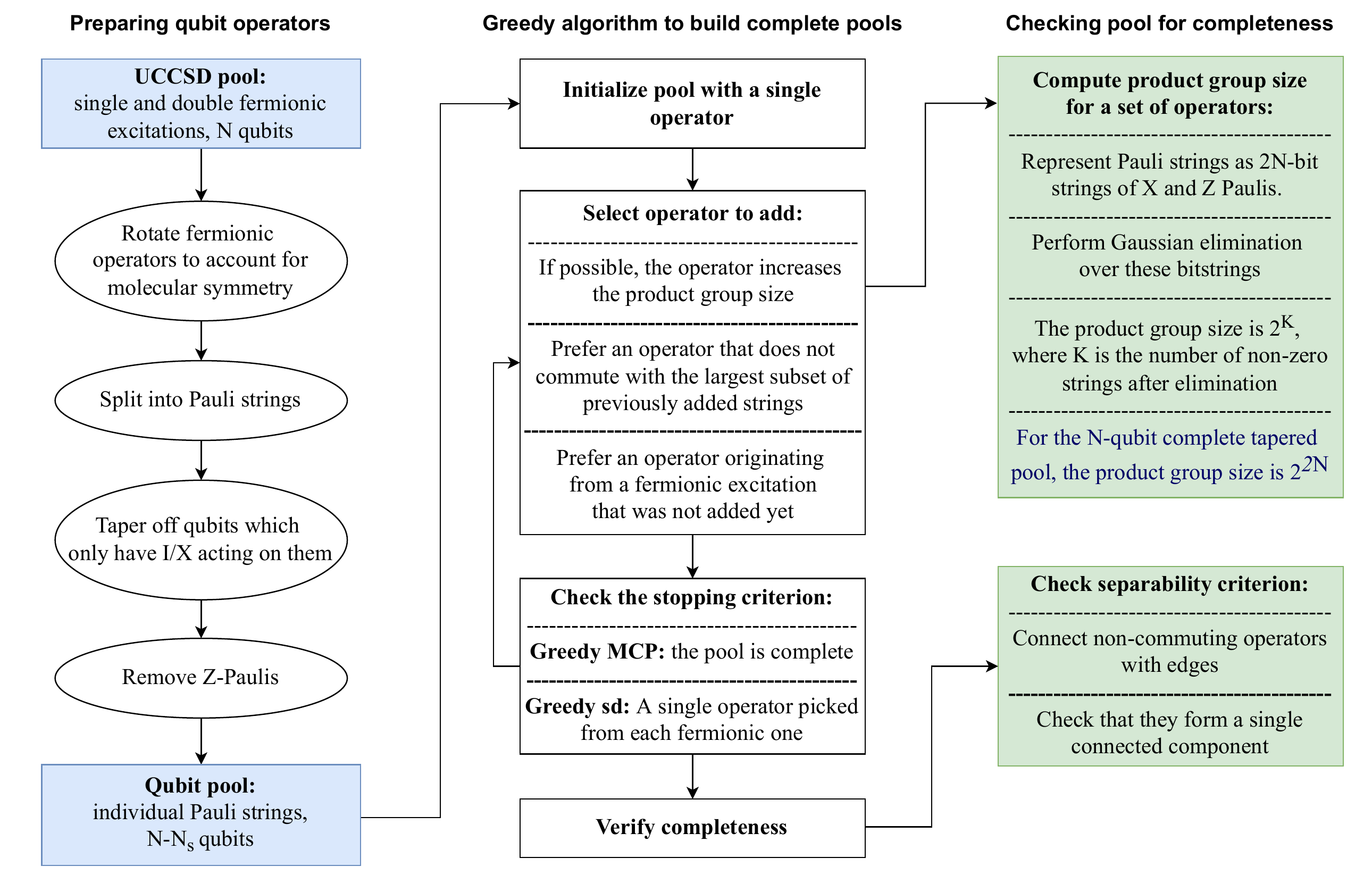}
    \centering
    \caption{\textbf{Computational scheme of complete qubit linear (Greedy MCP) and polynomial (Greedy sd) pools construction}.}
    \label{fig:mcp-algo}
\end{figure*}

Further study of MCPs by Shkolnikov et al.~\cite{shkolnikov2021avoiding} focused mainly on MCPs for molecular simulations. According to their results, the size of a minimal complete pool for molecular Hamiltonians is even less than for random Hamiltonians as it should preserve the {molecular} symmetries.
The authors of Ref.~\cite{shkolnikov2021avoiding} deeply analyzed these symmetries from the molecular group point of view and demonstrated how they force additional restrictions on the structure of Pauli string generators which form a complete pool. The following criteria have been formulated for minimal complete pools in molecular case~\cite{shkolnikov2021avoiding}:
\begin{enumerate}
    \item the number of electrons with a given spin changes by a multiple of 2;
    \item each operator in the pool must conserve spatial parity;
    \item the pool must contain enough ``starters'' for ADAPT-VQE to start; 
    \item the pool generates the biggest subgroup and subalgebra of those generated by a general non-symmetry-preserving MCP, that contain Pauli strings obeying conditions 1-2.
\end{enumerate}

The most important challenge in practice is {validating} of the fourth criterion: the complexity of building Lie algebra scales exponentially with the number of qubits. This criterion can be replaced by checking the product group size and the inseparability condition: MCPs generators in a complete pool cannot be split into two mutually commuting sets. Additionally, a difficult part is the pool construction since in the original work authors performed manual analysis of molecular symmetries.

The practical implementation of qubit ADAPT-VQE with MCPs requires simple and automated ways of building the complete pools. Here we demonstrate how to apply the proposed pool completeness criteria to the MCPs built in the reduced qubit space, \emph{i.e.}, with qubit tapering procedure (see Methods, \Cref{subsec:qubit_pool}). Thus, we not only build a pool of linear size for the molecule, but also remove qubits from the simulation, which significantly reduces the required computational resources.
Additionally, we propose a greedy algorithm for building complete molecular pools from a predefined large pool. In this work, we use qubit pools originated from tapered fermionic UCCSD pools.

The proofs for the complete pool construction combined with qubit tapering are presented in the Methods section. Here, we only highlight the main steps of building a complete pool. We propose to select Pauli strings from the set of tapered UCCSD operators instead of manually constructing them. First, the Pauli strings from UCCSD satisfy the complete pool conditions for the considered molecules (see Methods~\Cref{subsec:qubit_pool}). Second, most of Pauli strings from UCCSD correspond to double excitations that are starters for ADAPT-VQE, which allows {satisfying} criterion 3 without additional work. We propose the following scheme to build a complete pool in a reduced qubit space (a detailed scheme is illusrated in Fig.~\ref{fig:mcp-algo}).
\begin{enumerate}
    \item \textit{Taper off qubits based on Z$_2$ symmetries for both molecular hamiltonian and UCCSD pool}. This can be done using a built-in procedure in Qiskit or other frameworks for quantum computations. A more detailed explanation of the procedure can be found in Methods.
    \item \textit{Cut the tapered UCCSD operators into individual Pauli strings}. We additionally remove all Z-Paulis from the obtained strings. Although it is not a necessary step, it makes circuits significantly shallower according to our numerical investigation. At this step, we also check the completeness of the pool, which is inexpensive from the computational point of view (see Gauss method in Methods).
    \item \textit{Select $2(N-N_s)$ strings from the pool constructed in the previous step to form a complete pool.} Here $N_s$ is a number of tapered qubits. We propose a greedy approach to select the strings that form a complete pool. 
\end{enumerate}

The Pauli string selection from the UCCSD pool (step 3) can be implemented in various ways.
As mentioned above, to create a complete pool, we need to check the size of the product group and pool inseparability. The detailed instructions on checking the above conditions are given in the Methods. We use a greedy algorithm to construct a complete pool by iteratively adding strings to the pool one by one. We follow several rules to construct the pool: (i) {to} satisfy the group size criterion, we choose only the strings that increase the group size (\emph{i.e.}, linearly independent with already added strings); (ii) {to} meet the inseparability criterion, we select a string {that} does not commute with the largest subset of added strings; (iii) to make the operators more diverse, we try to add the Pauli strings, which act by $X$/$Y$ on different sets of qubits. Although the other ways of pool construction are possible, the numerical results confirm that the performance of the obtained greedy MCP pools is at least on par with the ones tested in the original work~\cite{shkolnikov2021avoiding} for H$_4$ and LiH molecule. The advantage of the proposed algorithm for obtaining MCP lies in the combination of qubit reduction and the simplicity of the pool construction: by using operators from tapered fermionic excitations, we account for all the necessary symmetries, while the greedy algorithm builds a pool of the required rank.

\begin{figure*}[htbp]
\centering
    \includegraphics[width=0.9\textwidth]{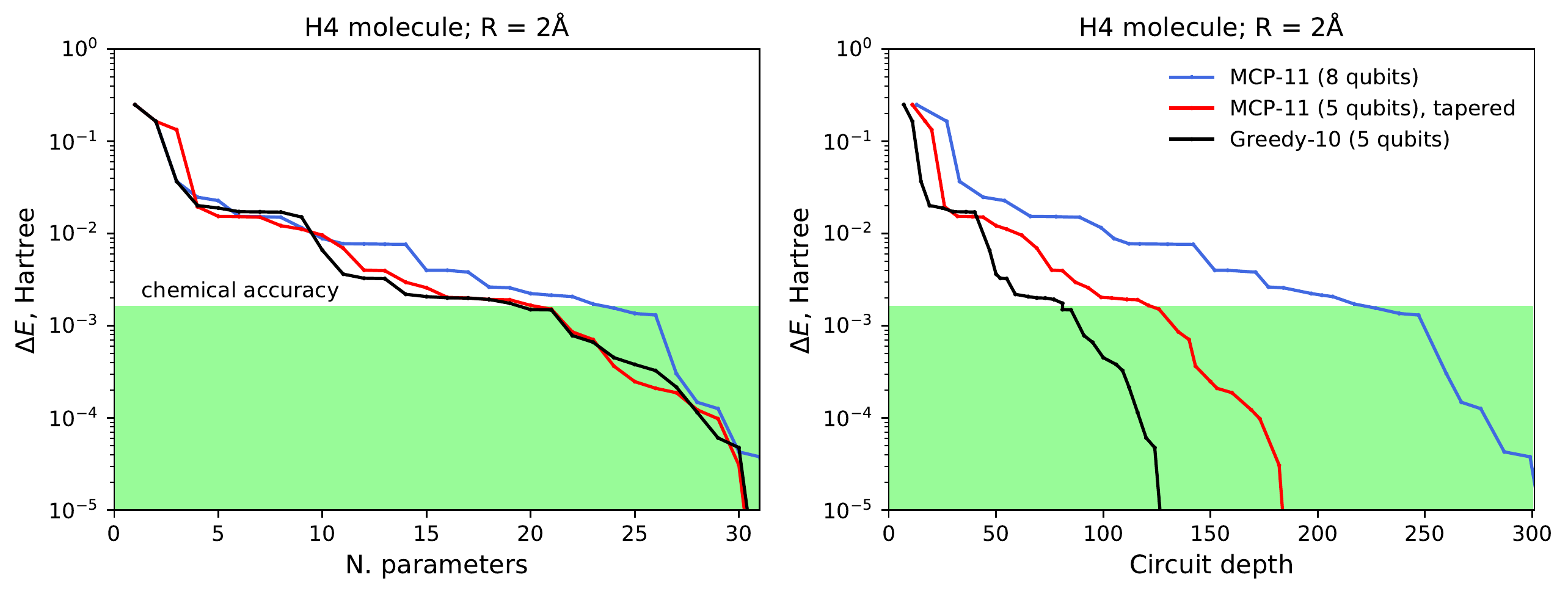}
    \centering
    \caption{\textbf{Qubit ADAPT-VQE convergence of energy with respect to the number of parameters and the circuit depth for H$_4$ molecule at $R = 2.0${\AA}.} MCP-11 is the pool acting in the original qubit space of 8 qubits suggested in Ref.~\cite{shkolnikov2021avoiding}. `MCP-11, tapered' is built from MCP-11 by applying tapering procedure. Greedy-10 pool is constructed using the computational scheme proposed in this work and acts in a 5-qubit space.}
    \label{fig:h4_adapt-vqe-conv}
\end{figure*}

We perform simulations for the H$_4$ molecule in the minimal basis set (STO-3G) with the MCP pool given in Ref.~\cite{shkolnikov2021avoiding} (see MCP-11 in Fig.~\ref{fig:h4_adapt-vqe-conv}) on 8 qubits for the bond length $R$ being equal to $1.0$ and $2.0${\AA}. We apply the tapering procedure to the given pool MCP-11: we rotate the molecular hamiltonian and the MCP generators with a unitary transformation, such that the rotated generators act only by X or I on a subset of qubits and eliminate these qubits. The obtained pool contains Pauli strings acting on 5 qubits (``MCP-11; tapered''). \Cref{fig:h4_adapt-vqe-conv} shows that the tapered pool converges to the ground state with high precision (results for $R = 1${\AA} are provided in Supplemental Information). {The} tapering procedure allows us to reduce the circuit depth significantly. We construct the ``greedy MCP'' pool consisting of 10 operators using the proposed computational scheme shown in Fig.~\ref{fig:mcp-algo}.
The greedy MCP pool demonstrates similar convergence with respect to the number of parameters compared to the tapered MCP-11, but reduces the required circuit depth due to the absence of $Z$ Paulis.

\begin{figure*}[htbp]
\centering
    \includegraphics[width=0.9\textwidth]{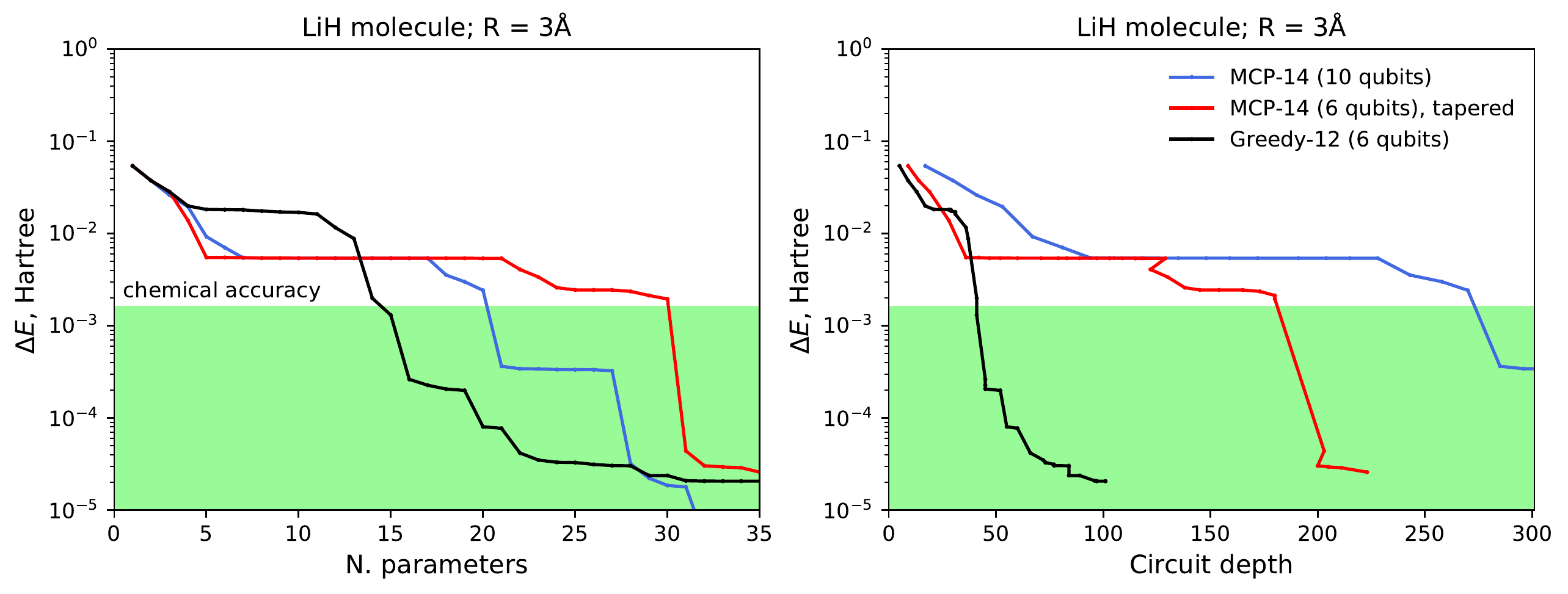}
    \centering
    \caption{\textbf{Qubit ADAPT-VQE convergence of energy with respect to the number of parameters and the circuit depth for LiH molecule at $R = 3.0${\AA}.} MCP-14 is the pool acting in the original qubit space of 10 qubits suggested in previous work~\cite{shkolnikov2021avoiding}. `MCP-14, tapered' is built from MCP-14 by applying tapering procedure. Greedy-12 pool is constructed using the computational scheme proposed in this work and acts in a 6-qubit space.}
    \label{fig:lih-convergence}
\end{figure*}

For the LiH molecule, after freezing the inner orbital, there are 2 electrons, which leads to only 10 excitations in VQE-UCCSD after symmetry reduction (see Methods,~\Cref{subsec:comp_details}). The small size of the tapered fermionic pool leads to a special case, where the Pauli strings from UCCSD pool without $Z$ Paulis do not form a complete pool. The rank of the pool is 10, while the required rank for a 6 qubit molecule should be 12. To overcome this difficulty, we first constructed the pool of 10 {operators} from the tapered UCCSD pool with removed Z paulis and then added two more operators with $Z$-paulis. The energy convergence with respect to the number of parameters and circuit depth for $R = 3${\AA} is given in \Cref{fig:lih-convergence}. As in the $H_4$ case, tapering allows us to reduce the required circuit depth. The constructed pool ``Greedy-12'' demonstrates the excellent convergence compared to tapered MCP-14 and reduces the circuit depth.

The results for H$_4$ and LiH are obtained at bond stretching when the correlation energy increases (the convergence of different pools near the equilibrium is provided in Supplementary Information).
Based on the provided numerical results, we expect that the proposed approach for pool construction is efficient for small molecules. Further, we use the pools constructed with the greedy algorithm to simulate larger molecules and compare the results to fermionic ADAPT-VQE approach and polynomial qubit pools.

\subsection{Simulation of molecules}\label{subsec:simulation}

Here we present the simulation results for a set of larger molecules. We perform simulations on the Qiskit statevector simulator~\cite{Qiskit} using various methods, including our improved version of the ADAPT-VQE approach -- batched ADAPT-VQE.
We start with benchmarking the methods on H$_2$O at various bond lengths. 
Further, we present the simulation results for molecules involved in CO-oxidation: O$_2$, CO, and CO$_2$. 

All the computations are performed in the minimal basis set using frozen core approximation and qubit tapering to reduce the required resources (see Methods, \Cref{subsec:comp_details}). 
We compare numerical results obtained with trotterized (or disentangled) VQE-UCCSD with SD ordering ~\cite{Grimsley_Trot}, ADAPT-VQE, and their various implementations (fermionic/qubit and original/batched).

To compare the computational cost, we calculate the total number of 1-parameter derivative computations $N_{d1}$ during the simulation: we consider both the ansatz growing phase in ADAPT-VQE case (\emph{i.e.}, derivatives with respect to each operator from the pool) and the VQE optimization procedure.

\subsubsection{H$_2$O molecule}\label{subsubsec: small_mols}

\begin{figure*}[htp]
    \centering
    \subfloat[Energy convergence with respect to the number of parameters for ADAPT-VQE and VQE-UCCSD.]{
        \includegraphics[width=\textwidth]{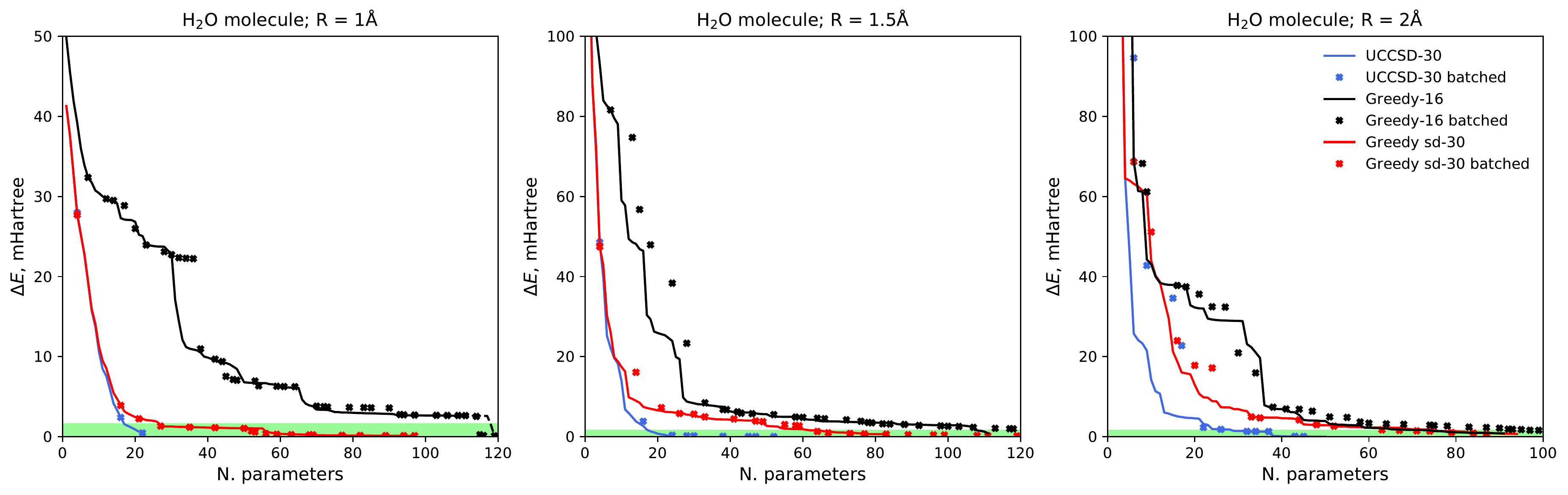}
        \label{fig:h2o-energy-parameters}
    }
    
    \subfloat[Energy convergence with respect to the total number of derivative evaluations $N_{d1}$ for ADAPT-VQE and VQE-UCCSD.]{
        \includegraphics[width=\textwidth]{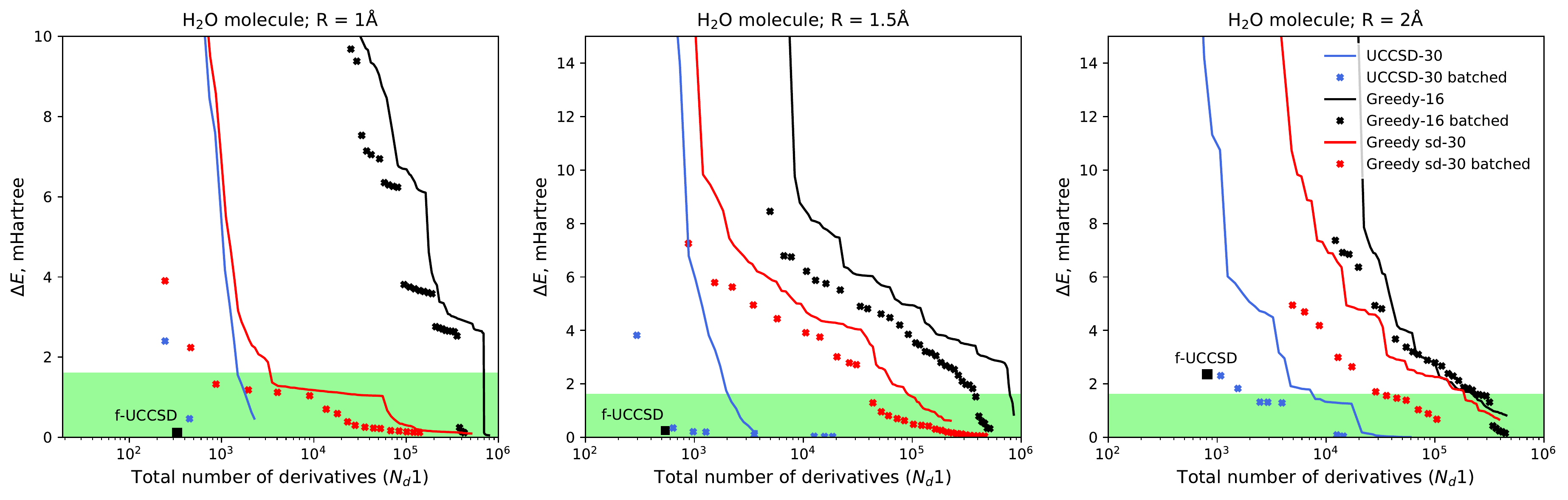}
        \label{fig:h2o-energy-derivatives}
    }
    
    \subfloat[Energy convergence with respect to the circuit depth for ADAPT-VQE and VQE-UCCSD.]{
        \includegraphics[width=\textwidth]{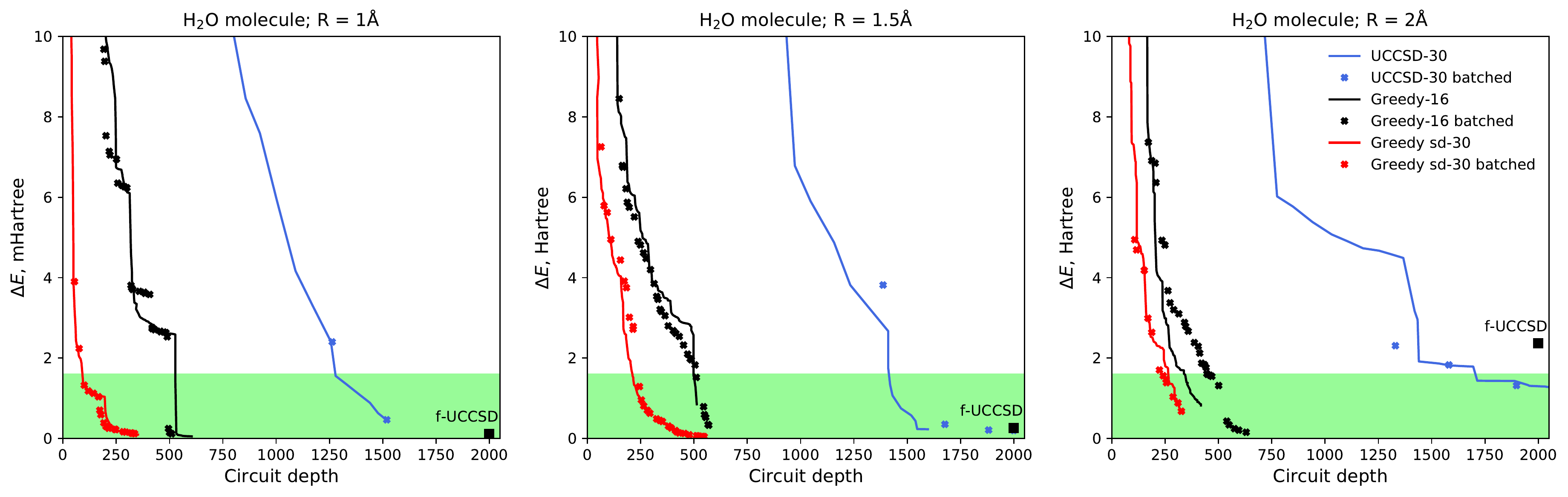}
        \label{fig:h2o-energy-depth}
    }
    
    \caption{\textbf{ADAPT-VQE and VQE-UCCSD convergence of the energy with respect to the number of parameters, total number of derivatives and circuit depth for H$_2$O molecule at various bond lengths.} Fermionic ADAPT-VQE results are marked as ``UCCSD-$N$'', which indicates the fermionic pool size for the molecule. Qubit ADAPT-VQE results are referred to as Greedy-$N$ (linear pool) and Greedy sd-$N$ (polynomial pool), where $N$ is the qubit pool size. With the green color we highlight the area, where the chemical accuracy threshold is satisfied.}
    \label{fig:h2o-convergence}
\end{figure*}

For H$_2$O, we simulate symmetrical stretching of two O-H bonds with a fixed angle of 104.51$^{\circ}$ (see Fig.~\ref{fig:h2o-convergence}). The UCCSD pool after symmetry reduction consists of 30 excitations acting on 8 qubits (see Methods, \Cref{tab:resources}). Close to the equilibrium geometry (bond length $R = 1.0${\AA}) and at $R = 1.5${\AA}, fermionic VQE-UCCSD provides accurate results for the ground state energy (see f-UCCSD in \Cref{fig:h2o-energy-derivatives},~\ref{fig:h2o-energy-depth}).

With the bond stretching and increase in the correlation energy at $R = 2.0${\AA}, VQE-UCCSD is not sufficient to reproduce the energy with the required accuracy. Fermionic ADAPT-VQE improves energy compared to VQE-UCCSD. At $R = 1.0${\AA} and $1.5${\AA} fermionic ADAPT-VQE reaches the chemical accuracy threshold with 17 and 18 parameters, respectively, and increases $N_{d1}$ more than in 4 times. As the H-O distance increases, the number of required parameters grows to 29 parameters at $R = 2.0${\AA}, which almost equals the size of the UCCSD pool. 
The total number of computed derivatives {is higher in this case by order of magnitude.}

Fermionic batched ADAPT-VQE reaches the threshold with a slight overhead in the number of parameters at these bond lengths --- 22, 24, and 32 --- which leads to an increase in the circuit depth by less than 20\% compared to the original ADAPT-VQE. At the same time, fermionic batched ADAPT-VQE requires at least 3 times fewer derivative computations ($N_{d1}$, derivatives with respect to a single parameter) in total for each bond length. It is worth noting that at $R = 2.0${\AA} batched ADAPT-VQE initially produces energies noticeably worse than the original implementation. {Still, over} a single iteration {it} improves the energy by 20 mHartree and at 22 parameters catches up with the original ADAPT-VQE values and further provides similar energies.

The observed results demonstrate that close to the VQE-UCCSD result, batched fermionic ADAPT-VQE is close to VQE-UCCSD in terms of computed derivatives $N_{d1}$, while the original ADAPT-VQE approache increases $N_{d1}$ significantly. Even though the convergence slows down near the threshold in the presence of strong correlation, batched fermionic ADAPT-VQE allows {reducing the} number of computed derivatives significantly.

To implement qubit ADAPT-VQE, we start by constructing a minimal complete pool with the proposed greedy approach. For H$_2$O we obtain the pool of 16 operators (Greedy-16), which includes no more than a single Pauli string from each initial fermionic excitation after removing $Z$ Paulis.
At $R = 1.0${\AA}, Greedy-16 demonstrates significantly slower convergence compared to fermionic ADAPT-VQE and reaches the energy threshold with 118 parameters. Batched ADAPT-VQE, in this case, reaches the threshold with 115 parameters and requires 1.8 times fewer derivatives. For $R = 1.5${\AA} the situation changes: original ADAPT-VQE requires 108 parameters to reach the threshold, while batched implementation produces a large overhead and converges at 123 parameters. However, from the \Cref{fig:h2o-energy-derivatives}, one can see that batched ADAPT-VQE still reduces the number of derivative computations by a half. Moreover, batched ADAPT-VQE does not lose in terms of the circuit depth here due to the usage of lightweight Pauli strings with no $Z$ chains as well as Qiskit's heavy transpilation of circuits.
Finally, at $R = 2.0${\AA} batched ADAPT-VQE performs worse than the original one and converges only with 101 parameters instead of 74 for original ADAPT-VQE and computing about 10\% more derivatives.

One can see that in comparison to fermionic ADAPT-VQE, qubit ADAPT-VQE with greedy MCP produces a significant overhead in the number of derivative computations required to obtain accurate results.
To verify that MCP in practice reduces the measurement overhead in qubit ADAPT-VQE, we compare the Greedy-16 performance with a qubit pool of size 30. This pool is obtained by the same greedy algorithm, but we pick 30 operators --- at least a single operator from each UCCSD excitation after removing $Z$ paulis --- instead of 16. We refer to such pools as ``Greedy sd-N'', where ``sd'' refers to picking operators from each UCCSD excitation. Thus, such pool is at least of the size of the UCCSD pool and satisfies the completeness criteria.

Our results demonstrate that such pool converges faster {than} MCP with respect to the number of parameters, total number of derivatives, and circuit depth. For $R = 1.0${\AA} qubit ADAPT-VQE with Greedy sd-30 pool converges at {a} significantly smaller number of parameters (27 for both original and batched) compared to $R = 2.0${\AA} (76 and 67 for original and batched, respectively), which matches the expectations as the correlation energy is small near the equilibrium.
These results indicate that the completeness criterion is {insufficient} to construct an efficient pool for qubit ADAPT-VQE computations. Batched ADAPT-VQE with Greedy sd pool performs similar to the original ADAPT-VQE in terms of the parameter number and circuit depth (see \Cref{fig:h2o-convergence}) while reducing the total number of derivatives $N_d1$ significantly: from 2.2 fewer derivative evaluations at $R = 1.5$ up to 5 times at $R = 2.0${\AA}. One can see that by using batched qubit ADAPT-VQE with Greedy sd pool we can reduce $N_{d1}$ by 1-2 orders of magnitude compared to {the} original qubit ADAPT-VQE with greedy MCP.

Considering our results, enlarging the qubit pool in the combination with batched ADAPT-VQE procedure appears to be {an} efficient implementation of qubit ADAPT-VQE. This observation contradicts the idea that the construction of a minimal linear pool {reduces} the measurement overhead. Numerically, it appears that limitation of the pool size leads to a significant increase in the number of iterations in the ansatz construction procedure.

\subsubsection{O$_2$ molecule}\label{subsubsec:o2_mol}

For the molecules involved in CO oxidation, we perform simulations at equilibrium geometries, precomputed classically with CCSD/STO-3G.
The ground state of the molecular oxygen is triplet, with two electrons remaining unpaired on the $\pi$ orbital. 
This is not a typical case for neutral molecules and thus a good testing case for ADAPT-VQE on open-shell systems. 
For O$_2$, we use the unrestricted Hartree-Fock (UHF) state as a reference state in ADAPT-VQE and VQE-UCCSD computations.

With the use of the qubit tapering procedure, the Hamiltonian can be reduced from 16 to 11 qubits, while the operator pool size becomes relatively small for O$_2$ molecule and consists of 24 excitations. 
VQE-UCCSD converges to the energy slightly above the chemical accuracy threshold (around 2 mHartree). Fermionic ADAPT-VQE (original and batched) converges to the fermionic VQE-UCCSD energy at 24 parameters, meaning no significant benefit in the parameter number observed in this case. 
Fermionic ADAPT-VQE reaches the chemical accuracy threshold with 27 parameters for both original and batched implementations. Batched ADAPT-VQE converges very similar to the original implementation producing similar quantum circuits and {reducing} the number of total derivative evaluations by 1.5 times at 27 parameters.

\begin{figure*}[htp]
    \centering
    \subfloat[ADAPT-VQE and VQE-UCCSD convergence of the energy with respect to the number of parameters, total number of derivatives and circuit depth for O$_2$ molecule. As far as single qubit operator, which is obtained from each fermionic excitations do not form a complete pool, we add two extra operators to ensure pool completeness. Thus greedy sd pool contains 26 operators in this case. ]{
        \includegraphics[width=\textwidth]{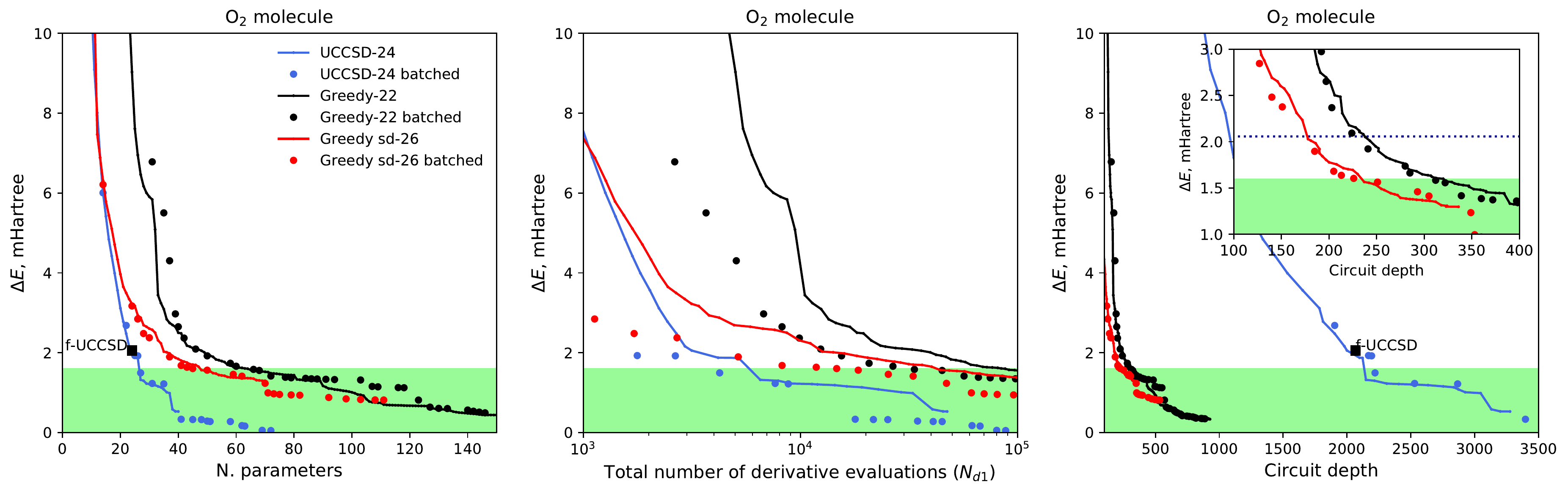}
        \label{fig:o2-convergence}
    }
    
    \subfloat[ADAPT-VQE and VQE-UCCSD convergence of the energy with respect to the number of parameters, total number of derivatives and circuit depth for CO molecule.]{
        \includegraphics[width=\textwidth]{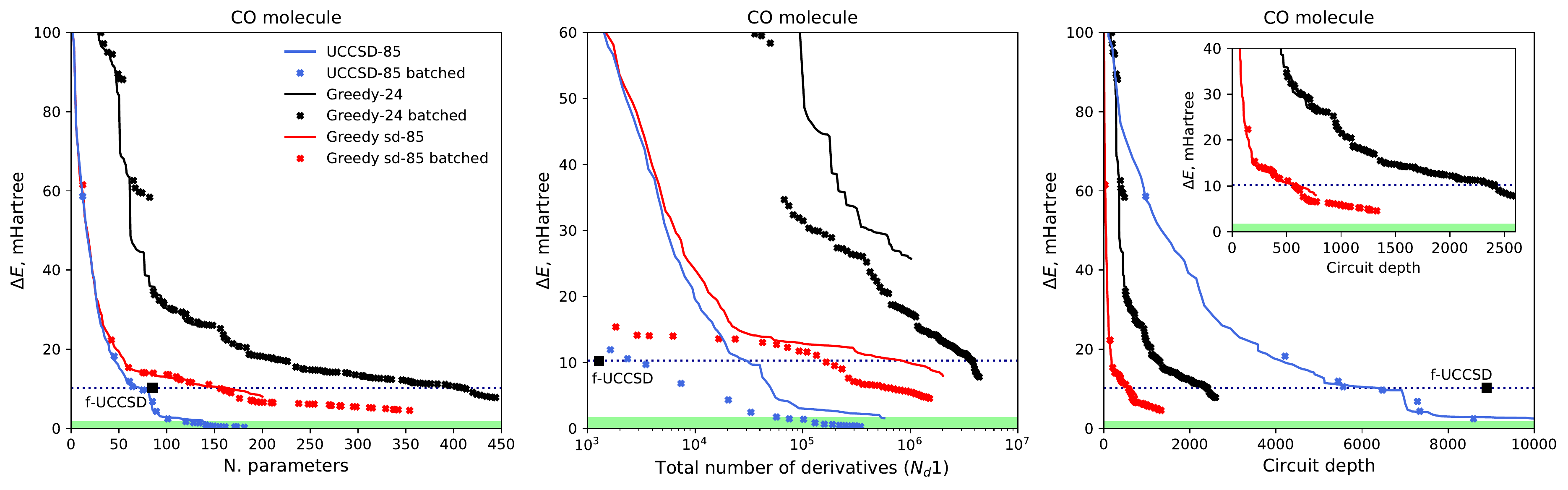}
        \label{fig:co-convergence}
    }
    
    \subfloat[ADAPT-VQE and VQE-UCCSD convergence of the energy with respect to the number of parameters, total number of derivatives and circuit depth for CO$_2$ molecule.]{
        \includegraphics[width=\textwidth]{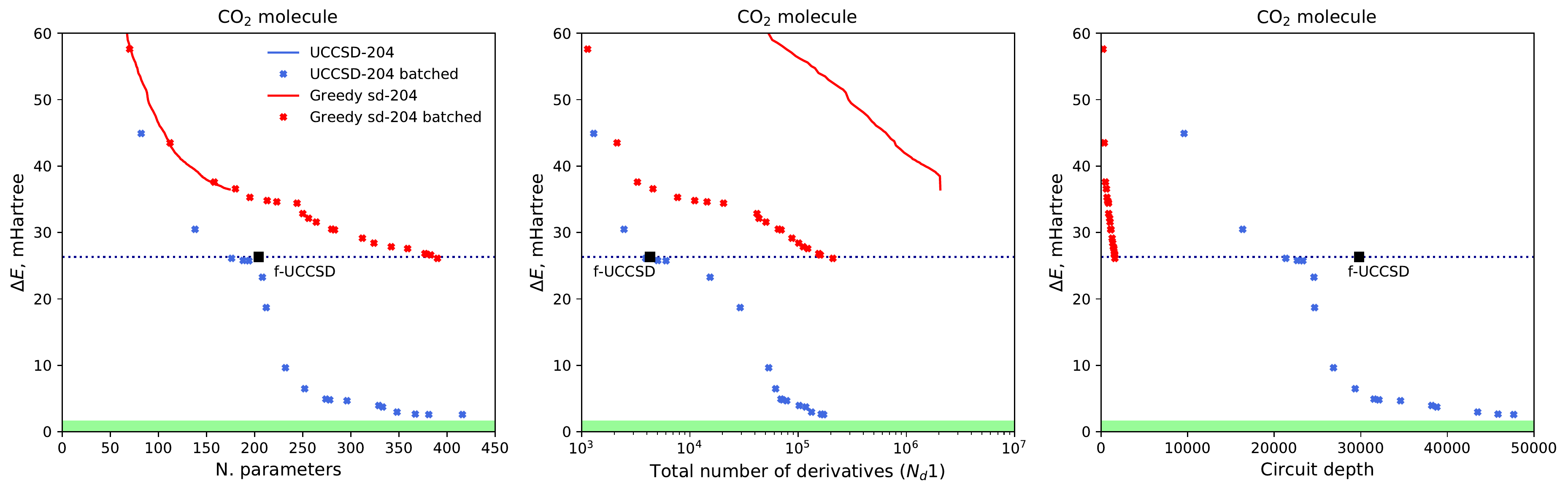}
        \label{fig:co2-convergence}
    }
    
    \caption{\textbf{ADAPT-VQE and VQE-UCCSD convergence of the energy with respect to the number of parameters, the total number of derivatives and the circuit depth for O$_2$, CO and CO$_2$ molecules computed at the equilibrium geometry.} Fermionic ADAPT-VQE results are marked as ``UCCSD-$N$'', which indicates the fermionic pool size for the molecule. Qubit ADAPT-VQE results are referred to as Greedy-$N$ (linear pool) and Greedy sd-$N$ (polynomial pool), where $N$ is the qubit pool size. By the green color we highlight the area, where the chemical accuracy threshold is satisfies.}
    \label{fig:o2-co-co2-convergence}
\end{figure*}

Greedy MCP for O$_2$ requires 22 operators (Greedy-22) and reaches the chemical accuracy threshold with 65 and 66 {parameters}  for original and batched ADAPT-VQE, respectively, which reduces the circuit depth from about 2,200 in fermionic ADAPT-VQE to 300. Batched ADAPT-VQE requires 2.7 times less derivative evaluations than {the} original one. 
Due to the small size of the tapered UCCSD pool we {cannot} pick a single operator from each excitation (after removing Z Paulis) and form a complete pool from such operators. For this reason, we add two extra operators to ensure the pool completeness. The obtained pool contains 26 operators (Greedy sd-26), which is very close to the MCP size. Qubit ADAPT-VQE converges slowly for both Greedy-22 and Greedy sd-26. Nevertheless, adding a few operators to the pool can improve the convergence in terms of both the circuit depth and {the} total number of derivatives. Batched qubit ADAPT-VQE with Greedy sd-26 pool performs better than other qubit ADAPT-VQE implementations in this case. Comparing two extreme cases -- batched ADAPT-VQE in combination with Greedy sd-26 pool and original ADAPT-VQE with MCP pool -- we can reduce the number of total derivative evaluations by 5 times.
It confirms our idea that decreasing the pool size is not sufficient to reduce the number of required measurements. On the contrary, O$_2$ results follow the pattern observed for H$_2$O {for} a larger operator pool: the convergence of qubit ADAPT-VQE improves while the computational cost reduces.

\subsubsection{CO molecule}\label{subsubsec:co_mol}

Carbon monoxide is a more complicated case compared to oxygen due to the larger number of electronic excitations. In computational chemistry, carbon monoxide is a known example when density functional theory (DFT) fails, which is appeared in the incorrect prediction of the CO adsorption on the transition metal surface~\cite{Soini_2015, Feibelman_2001}.
The observed inconsistencies between classical simulations and experimental results make quantum {simulations} of reactions involving carbon monoxide an attractive problem with {the} proper development of quantum technologies. 
At this stage, we are only able to simulate an isolated CO molecule. 
However, even in the minimal basis set, it represents an interesting test case for quantum methods.

VQE-UCCSD simulations of CO molecule with STO-3G basis set have been reported previously~\cite{armaos2019computational}. 
This simulation uses 20 qubits since the symmetry reduction and the orbital freezing are not applied. 
The reported VQE-UCCSD energy (-111.363 Hartree) matches the classical CCSD result (-111.362 Hartree), which is about 10 mHartree higher than the exact one. 
This is a sizable deviation compared to small molecules, where VQE-UCCSD at equilibrium geometry accurately recovers correlation energy. 
Our simulations of CO require only 12 qubits, which greatly reduces the computational cost. 
VQE-UCCSD with 12 qubits for CO molecule fits previously reported results. 

ADAPT-VQE convergence to the ground energy is presented in \Cref{fig:co-convergence}. 
Fermionic ADAPT-VQE reaches VQE-UCCSD energy with 72 and 75 parameters (original and batched implementations), which is significantly less than the UCCSD pool size (85 operators). This allows {reducing} the circuit depth by about 30\% compared to VQE-UCCSD. Fermionic batched ADAPT-VQE requires about 9 times fewer derivatives to reach VQE-UCCSD energy compared to the original one. When reaching the chemical accuracy threshold, batched ADAPT-VQE performs better as well: it converges with 128 parameters instead of 134 and with about 7 times fewer derivative computations.

Thus, fermionic batched ADAPT-VQE rapidly converges to VQE-UCCSD energy and has a reasonable overhead in the number of derivative evaluations (about 3,500 $N_{d1}$ for batched fermionic ADAPT-VQE compared to 1,275 $N_{d1}$ for fermionic VQE-UCCSD), while original ADAPT-VQE requires about 30,100. The observed results match the ones obtained for H$_2$O: batched implementation of fermionic ADAPT-VQE can achieve VQE-UCCSD results more efficiently than VQE-UCCSD while significantly reducing the computational cost compared to original ADAPT-VQE. However, beyond this point the further energy improvement becomes slow from the optimization point of view, with the number of computed derivatives {increasing} to around 75,000 for batched ADAPT-VQE and 555,000 for original ADAPT-VQE. The original fermionic ADAPT-VQE produces {a} huge overhead in the number of derivative evaluations.

We {perform} qubit ADAPT-VQE computations for Greedy and Greedy sd pools for the CO molecule. Greedy MCP contains 24 operators (Greedy-24) for 12-qubit CO molecule, while the Greedy sd pool consists of 85 Pauli strings (Greedy sd-85). For Greedy-24 we perform original ADAPT-VQE computations up to 140 parameters to see if batched ADAPT-VQE matches its convergence curve.

With batched qubit ADAPT-VQE, we do not achieve chemical accuracy due to the high computational cost, but we achieve the level of the fermionic VQE-UCCSD energy. Both pools allow {reducing} the circuit depth significantly compared to UCCSD. Qubit ADAPT-VQE with Greedy sd-85 pool achieves VQE-UCCSD result with 164 and 156 parameters with original and batched implementations, respectively. Batched ADAPT-VQE reduces the number of computed derivatives by a factor of 5. Qubit ADAPT-VQE with Greedy sd pool turns out to be the most efficient in terms of circuit depth. However, the price we pay for the circuit reduction is the increase in the number of derivative evaluations by two orders of magnitude compared to VQE-UCCSD.

At the same time, batched ADAPT-VQE with Greedy-24 requires 413 parameters to reach VQE-UCCSD energy. The number of derivative evaluations is about 23 times higher for the MCP pool compared to Greedy sd-85 when computed with batched ADAPT-VQE. 

The observed results demonstrate a vastly worse convergence of greedy pool than greedy sd, which again confirms that {in practice, the linear pool does not reduce the measurement overhead, especially with the increase of the molecule complexity and size.}

\subsubsection{CO$_2$ molecule}\label{subsubsec:co2_mol}

According to precise classical \textit{ab initio} calculations, 
an accurate description of reaction energies involving carbon dioxide requires incorporating triple and even higher order of excitations in coupled cluster ansatz~\cite{Douglas_Gallardo_2019}. Such a slow convergence arises from the two degenerated $\pi$-molecular orbitals.

The full simulation of carbon dioxide in STO-3G basis set requires 30 qubits, which can be further reduced to 19 by freezing core orbitals and tapering 5 qubits considering the symmetry point group. 
To the best of our knowledge, it is the first such {sizeable} numerical computation of CO$_2$ ground state energy using quantum computing devcies.

Due to the complex correlation effects, CCSD significantly deviates from the exact ground state energy at the equilibrium geometry. 
Energy recovered by fermionic VQE-UCCSD for carbon dioxide is close to the classical CCSD and is about 25 mHartree higher than the exact energy. 

For fermionic ADAPT-VQE, we perform 55 iterations to check if batched ADAPT-VQE fits the original convergence curve (see \Cref{fig:co-convergence}). 
Based on the previous results, we expect batched ADAPT-VQE to perform similarly to original ADAPT-VQE in circuit size.
Fermionic batched ADAPT-VQE requires only 4 iterations and 176 parameters to achieve fermionic VQE-UCCSD energy. The number of parameters is thus significantly reduced compared to UCCSD, which has 204 excitations.
The number of computed derivatives in batched fermionic ADAPT-VQE procedure is almost the same as in VQE-UCCSD (3,900 and 4,300 $N_{d1}$) in this case, due to the great parameter savings. 
As it is observed for other molecules, the convergence slows down significantly when we try to improve energy beyond VQE-UCCSD and approach the exact result.
We perform simulations up to doubling the parameter number compared to the UCCSD pool size for fermionic and qubit batched ADAPT-VQE. 
For fermionic batched ADAPT-VQE, we {achieved} energy about 2.5 mHartree higher than the FCI result, which is very close to the chemical accuracy threshold. 

Since greedy MCP is not efficient for H$_2$O and CO molecules, we run the simulation only for the Greedy sd pool due to the high computational cost of CO$_2$ simulation. For original qubit ADAPT-VQE, we perform around 175 iterations to reach the gradient norm less than $10^{-1}$ (see Methods, \ref{subsec:convergence_criteria}). 
The energy difference between batched and original qubit ADAPT-VQE is less than 1 mHartree at {a} fixed number of parameters {confirming} batched ADAPT-VQE's decent performance.
Batched qubit ADAPT-VQE approaches VQE-UCCSD with 390 parameters (see \Cref{fig:co2-convergence}) and the circuit depth of 1,560 in contrast to 24,800 for the fermionic VQE-UCCSD. 
{However, such circuit reduction goes at the price of 40 times more derivative computations.} 

\subsection{Electronic energy of reaction}\label{subsec:el_energy}

Since relative energies but not their absolute values are crucial to determine the accuracy of \textit{ab initio} methods, we can compare the electronic energy of CO-oxidation reaction computed by the studied methods:
\begin{equation}
    \text{CO} + \frac{1}{2}\text{O}_2 \rightarrow \text{CO}_2.
\end{equation}
Electronic energy of this reaction is given by the following formula:
\begin{equation}
    \Delta E_r = E_{{\rm CO}_2} - E_{CO} - \frac{1}{2}E_{{\rm O}_2}
\end{equation}

\Cref{tab:reaction-energy} shows the electronic energy of reaction computed by different methods. 
Classical CCSD only slightly improves the uncorrelated Hartree Fock result. 
{The} inclusion of triplet excitations in coupled cluster  is necessary {for accurate estimation} of the electronic reaction energy.

We compute {the} electronic energy of the reaction using only the batched implementation of ADAPT-VQE
because CO$_2$ molecule is too computationally expensive for the original ADAPT-VQE simulation. We compare energies obtained with different stopping criteria. 
By setting the maximum derivative value to $10^{-3}$ (or alternatively {the} gradient norm to $10^{-2}$) we could obtain accurate results for batched fermionic ADAPT-VQE. 
When stopping the simulation after reaching $\max{g_i} \le 10^{-2}$, batched ADAPT-VQE obtains energies close to VQE-UCCSD, but with shallower circuits.
Therefore, the provided results for fermionic batched ADAPT-VQE can be regarded as a theoretical benchmark of the adaptive ansatz for reaching FCI.

Qubit batched ADAPT-VQE converges significantly slower, which leads to a poor description of the reaction's electronic energy. It is worth noting that energy underestimation also relates to the fact that qubit ADAPT-VQE quickly converges for O$_2$ molecule compared to CO and CO$_2$. 
It should be noted that both the maximum derivative value and gradient change non-monotonically, {leading} to a significant variance in the estimated reaction energy. 
As we do not achieve $\max{g_i} \le 10^{-3}$ for CO$_2$, we estimate the ``best'' energy by taking the results for O$_2$ and CO at the number of parameters two times larger than in the UCCSD pool. 
The best {achieved} energy for qubit ADAPT-VQE only slightly improves HF result. 
Improving VQE-procedure is necessary for qubit ADAPT-VQE to speed up the convergence and obtain an accurate result. 

From the energies obtained with various stopping criteria, one can see sluggish convergence of qubit ADAPT-VQE on molecules with complicated electronic {structures}. 
Even stricter convergence criteria {are required to further energy improvement}, which would lead to a significant increase in the computational cost.
{Our simulation could} reach fermionic VQE-UCCSD energy, which is already a promising result, as we tested it on complex and relatively large molecules.

\begin{table}[]
\caption{\textbf{Electronic energy of CO oxidation reaction computed with the studied classical and quantum methods.} ADAPT results refer to simulations with batched ADAPT-VQE since the original ADAPT-VQE is too costly for CO$_2$ molecule. For ADAPT methods, we provide results after satisfying different convergence criteria.}
\begin{tabular}{ccc}
                         & \begin{tabular}[c]{@{}c@{}}Convergence\\ criterion\end{tabular} & $\Delta E_R$, kcal/mol \\ \hline
HF                       & -                                                               & -18.5                  \\
CCSD                     & -                                                               & -20.5                  \\
CCSD(T)                  & -                                                               & -30.7                  \\
FCI                      & -                                                               & -29.0                  \\ \hline
f-UCCSD                  & -                                                               & -19.5                  \\ \hline
\multirow{5}{*}{q-ADAPT} & best                                                       & -19.1                  \\
                         & $\max{g_i} \le 10^{-1}$                                         & -8.8                  \\
                         & $\max{g_i} \le 10^{-2}$                                         & -16.5                  \\
                         & $||g|| \le 10^{-1}$                                             & -16.0                  \\
                         & $||g|| \le 10^{-2}$                                             & -16.9                  \\ \hline
\multirow{5}{*}{f-ADAPT}
                         & $\max{g_i} \le 10^{-1}$                                         & -14.1                   \\
                         & $\max{g_i} \le 10^{-2}$                                         & -19.5                  \\
                      & $\max{g_i} \le 10^{-3}$                                         & -28.3                  \\
                         & $||g|| \le 10^{-1}$                                             & -20.1                  \\
                         & $||g|| \le 10^{-2}$                                             & -28.1                
\end{tabular}
\label{tab:reaction-energy}
\end{table}

\section{Discussion}\label{sec:discussion}

In this work, we have analyzed current implementations of the ADAPT-VQE approach for scalable quantum simulations. We have 
{simulated}
molecules that are of interest for accurate quantum chemical calculations. 
We have carried out simulations of bond stretching of H$_2$O molecule
{along with the} molecules with multiple strong bonds (O$_2$, CO, and CO$_2$). 

We have implemented fermionic and qubit ADAPT-VQE and proposed the algorithm implementation ---  batched ADAPT-VQE --- which adds {the} varying number of operators from the pool at each iteration. 
Such implementation allows reducing the number of derivate evaluations and speed up the ansatz growing procedure. 
The convergence of batched ADAPT-VQE confirms its efficiency in constructing a relatively compact ansatz for fermionic and qubit polynomial pools while demonstrating up to an order reduction in the total number of 1-parameter derivative evaluations. {The} adaptive size of the batch eliminates the computationally expensive procedure of selecting the appropriate pool size for each molecule separately.

As expected, with fermionic ADAPT-VQE it is possible to achieve VQE-UCCSD results with {the} reduced number of parameters and shallower circuits. {Still}, the parameter reduction in our case is not as impressive as it was reported by Grimsley et al.~\cite{Grimsley}. Besides the complex electronic structure of the considered molecules, the operator pools are reduced significantly by accounting for molecular symmetry (see \Cref{tab:resources}).

Our numerical results have demonstrated that batched fermionic ADAPT-VQE is a useful tool to improve the efficiency of VQE-UCCSD in terms of circuit depth while saving the number of computed derivatives significantly compared to {the} original fermionic ADAPT-VQE.

However, as the use of fermionic ADAPT-VQE is currently challenging for existing devices due to the deep quantum circuits, we pay {particular} attention to qubit ADAPT-VQE. 
Building on top of the work on minimal complete pools by Shkolnikov et al.~\cite{shkolnikov2021avoiding}, we combined MCP construction with qubit tapering procedure. Moreover, we proposed a greedy algorithm for automated pool construction from UCCSD operators that {do} not require manual analysis of molecular hamiltonian symmetries.  Our approach reduces the complexity of computations by reducing the number of qubits in the system. The constructed pools {demonstrate} excellent convergence for small molecules (H$_4$ and LiH), however, with the increase of the size and complexity of the molecules, we observe that qubit ADAPT-VQE with MCP produces a vast overhead in the number of computed derivatives. 
Along with the MCPs, we studied {polynomial-size qubit pools containing}
a single Pauli string from each UCCSD excitation after $Z$-pauli removal. Our numerical investigation leads to {a} suprising result: although theoretically, MCP is expected to reduce the measurement overhead in ADAPT-VQE, in practice, the qubit pool of the polynomial size requires significantly fewer derivative evaluations due to faster convergence. Moreover, the polynomial pool generates shallower circuits compared to MCP. Based on our numerical investigation, we suggest that {using} of batched ADAPT-VQE in combination with polynomial qubit pool is more efficient than reducing in the pool size in practice.

With qubit batched ADAPT-VQE using polynomial pools we {could} reach fermionic VQE-UCCSD results for all the considered molecules. Further improvement in the energy is possible, although {it} requires significant computational resources, namely, {an} increase in the number of computed derivatives due to the sluggish convergence. {Similar behavior} is also observed for the fermionic ADAPT-VQE in the case of a strong correlation when reaching the accurate energies. The bottleneck, in this case, lies in the VQE procedure, where we update all parameters simultaneously. Thus, we believe further study and improvements of the classical optimization routine are required to make progress and {enable the} practical application of ADAPT-VQE.

Despite the given shortcomings with {the} proper improvement of the optimization process, we expect both qubit and fermionic batched ADAPT-VQE {are promising for molecular simulation} due to the variable circuit size on near-term quantum devices. 
The investigation of ADAPT-VQE approaches in noisy conditions for moving towards experiments {on} real devices can be a subject of subsequent study.

\section{Methods}\label{sec:methods}

\subsection{Fermionic pool}\label{subsec:op_pool}

We start with the {spin-conserving} fermionic operators from the conventional UCCSD:
\begin{equation}\label{eq2}
    T(\theta) = \sum_{\substack{a \in \text{virt}\\i \in \text{occ}}}\theta_i^a a_a^\dag a_i + \sum_{\substack{a, b \in \text{virt}\\i, j \in \text{occ}}} \theta_{ij}^{a b} a_a^\dag a_b^\dag a_i a_j,
\end{equation}
where $a^\dag, a$ are fermionic creation and annihilation operators, respectively.
The summation is carried over occupied and virtual spin-orbitals.

After mapping excitations to qubits, the fermionic excitations are rotated following symmetry transformation~\cite{Setia_2020} such that the operators in qubit representation act by $X$ or $I$ on the subset of qubits. 
These qubits are further removed from the simulation. 
Thus, after {the} qubit tapering procedure fermionic operators may include{an} odd total number of $X$ and $Y$ Pauli operators.

\subsection{Qubit pool completeness}\label{subsec:qubit_pool}

According to the theorems of {completeness}~\cite{shkolnikov2021avoiding}, a minimal complete pool must contain $2N-2$ odd Pauli strings (\emph{i.e.}, strings having {an} odd number of $Y$ Paulis), which generate the product group $G$ with $\frac{2^{N-1}(2^{N-1} + 1)}{2}$ odd Pauli strings. Let us leave out of the bracket the minimality of the pool and consider a pool with $2N$ generators that {span} all possible $N$-qubit Pauli strings, $4^N$ in total. 
{Adding} back the two operators, we thus lose the information about the connection of two qubits, which was approved by Tang et al.~\cite{Tang_2021}. Although we do not have an analytical proof for {the} general case, for the considered molecules, we observed that this connection forms between tapered qubits and thus {could} be skipped out. Moreover, the pool size increases only by a constant, which does not affect the complexity. bserved on the considered molecules.

Consider an $N$ qubit molecule and a set of $2N$ odd Pauli strings, which generates a full product group with $4^N$ possible Pauli strings ($N$ positions with one of 4 matrices $I, X, Y$ and $Z$ at each). The number of odd Pauli strings in the full group is $2^N(2^N-1)/2$. The Lie algebra for the given set of operators consists of odd Pauli strings only, and its size is equal to $2^N(2^N-1)/2$ when the set forms a complete pool. Our goal is to find the size of the Lie algebra for the case of {the} symmetric molecular hamiltonian.

The unitary transformation $U$, which acts on the reference state $U|\text{Ref}\rangle = |\Psi\rangle$ is a superposition of operators -- Pauli strings -- from the Lie algebra, and all the operators must obey the restrictions raised from the hamiltonian symmetries. This means that {the} operator pool does not need to generate the full Lie algebra, only its subalgebra with all ``symmetric'' odd Pauli strings. Based on $Z_2$ symmetries, we can easily find the symmetry restrictions.
For the qubit molecular hamiltonian, there exist $N_{sym}$ Pauli strings ${S}$, which commute with the hamiltonian and {are} called symmetries. Based on these symmetries a unitary operator can be found that rotates a hamiltonian in such a way that it acts trivially or with at most one Pauli-gate (e.g. $I/X$) on a set of qubits $N_s$.

Now consider a subgroup of the full product $G$ group, which spans all the symmetric odd Pauli strings $G_{os}$ (we generate the full group and then select only the odd strings that commute with Z2 symmetries).
If we rotate all strings from $G_{os}$ by the unitary transformation $U_R$, the rotated strings $U_RG_{os}U_R$ have a specific structure and act by I or X on a subset of qubits.

The substring acting on $N_s$ qubits is always even for the rotated strings at it contains no Y Paulis on these positions. The number of odd Pauli strings then have the size of Lie algebra built on $N-N_s$ qubits multiplied by the number of possible combinations on the $N_s$ qubits: 
\begin{equation}
    \frac{2^{N-N_s}(2^{N-N_s} - 1)}{2}2^{N_s}.
\end{equation}
If we taper off $N_s$ from the computation, then the operator strings are reduced to the length $N-N_s$. The Lie algebra after tapering includes
\begin{equation}
    \frac{2^{N-N_s}(2^{N-N_s} - 1)}{2}
\end{equation}
elements, and the product group has $2^{N-N_s}$ pauli strings in total.

From the practical point of view, the considerations presented above allow us to formulate the completeness criterion for the molecule after the tapering procedure. {Suppose} the operator pool of odd Pauli strings generates a product group of the size $2^{N-N_s}$ and cannot be separated into two mutually commuting sets. {In that case,} the pool is complete for $N-N_s$ qubit molecular hamiltonian.

To check the completeness criteria, we use the following procedure. To compute the product group size. we represent individual $N$-qubit Pauli strings as bitstrings of length $2N$. This way, each Pauli matrix is represented by two bits -- $Z$ and $X$. $I$ matrix corresponds to zeros in both bits, while $Y$ corresponds to both ones. For example, the Pauli string $IXZY$ transforms into the following bitstring:
\begin{equation}
IXYZ  \longrightarrow (00)(01)(11)(10) \longrightarrow (00011110).
\end{equation}
{This} representation product of two Pauli strings is equivalent to the sum modulo two of their bitstrings up to a coefficient. 
For example:
\begin{equation}
\begin{split}
IXY \cdot ZZX \longrightarrow (000111) \oplus (101001) = \\
= (101110) \longrightarrow c \cdot ZYZ.
\end{split}
\end{equation}

To find the size of the product group for a certain pool, we need to know the size of the span for the {corresponding} bitstrings. 
We do so by running Gaussian elimination and finding {the} size $K$ of the minimal generating set. 
The size of the product group is then equal to $2^K$.

To check the inseparability criterion, we build a graph with vertices representing Pauli strings from the pool. 
We connect vertices with an edge {if} the Pauli strings do not commute. 
The criterion is satisfied when the constructed graph has a single connected component.
Both checks run in $\mathcal{O}(M^2N)$ time, where $M$ is the pool size, which is efficient enough.

\subsection{Computational details}\label{subsec:comp_details}

In this work, we perform noiseless (ideal) statevector simulations for the following set of molecules: H$_4$, LiH, H$_2$O, O$_2$, CO, and CO$_2$ in the minimal basis set (STO-3G). 
{We use both frozen core approximation and molecular symmetry for all the molecules to optimize the required resources: both number of qubits and number of electronic excitations.}
To map the ansatz to a quantum circuit, we use Jordan-Wigner transformation~\cite{JordanberDP}. 
Accounting for the molecular symmetry allows us to taper up to 5 qubits for the given molecules and reduce the operator pool significantly by excluding excitations ={that} commute with the qubit Hamiltonian (see \Cref{tab:resources}). 
The symmetry reduction leads to a great decrease in the computational time on a classical simulator. Although circuit complexity in VQE-UCCSD can be reduced by excluding excitations based on the precomputed MP2 amplitudes~\cite{Romero_2018, Kuhn}, we do not use any specific techniques suitable only for one of the methods.
VQE results are provided for trotterized (or disentangled) UCCSD with SD ordering (singles applied first to the reference state)~\cite{Grimsley_Trot}.

\begin{table}[]
    \caption{\textbf{Estimated resources for UCCSD ansatz with and without symmetry reduction.} 
    Columns represent the number of qubits in the molecular Hamiltonian, number of single excitations, and number of double excitations. 
    All the values are provided for the minimal basis set within frozen core approximation.}
    \centering
    \begin{tabular}{ ccccccc } 
         & N. qubits & N. singles & N.doubles \\
        \hline
        H$_4$ (no symm) &  8 & 8 & 18\\
        H$_4$ (symm) & 5 & 4 & 10\\
        \hline
        LiH (no symm) &  10 & 8 & 16\\
        LiH (symm) & 6 & 4& 6\\
        \hline
        H$_2$O (no symm) & 12 & 16 & 76\\
        H$_2$O (symm) & 8 & 6 & 24\\
        \hline
        CO (no symm) & 16 & 30 & 285 \\ 
        CO (symm) & 12 & 10 & 75 \\
        \hline
        O$_2$ (no symm) & 16 & 22 & 135 \\ 
        O$_2$ (symm) & 11 & 2 & 22 \\
        \hline
        CO$_2$ (no symm) & 24 & 64 & 1360 \\ 
        CO$_2$ (symm) & 19 & 12 & 192 \\
    \end{tabular}
    \label{tab:resources}
\end{table}

We use Qiskit library~\cite{Qiskit} for statevector simulations of quantum algorithms. All the circuit dimensions are given after {applying built-in ``heavy'' transpilation in Qiskit. The circuit depth is given by the number of consecutive one-qubit gates and CNOTs.} Even though more efficient circuits for implementing fermionic and qubit excitations have been proposed~\cite{Yordanov_fermionic, Yordanoc_iqeb}, they can not be applied directly in our case due to the tapering procedure.
We compare the results to the classical methods of quantum chemistry, such as CCSD, CCSD(T), and FCI (exact result in the given basis set). 
All the classical results are computed with PySCF package for quantum chemistry~\cite{Sun_2020}.

\subsection{Optimization strategy}\label{subsec:opt_strategy}

The classical optimization routine in the VQE approach can be implemented in several ways. 
We initialize all parameters for VQE-UCCSD with zeros. 
For the ADAPT-VQE, we set initial parameters to optimal values obtained at the previous iteration, as was done in the original work~\cite{Grimsley}. 
Such initialization improves classical optimization, as parameters are already optimized to a quite sensitive level.

Gradient-based methods {are} considered to be efficient in statevector simulation for both VQE and ADAPT-VQE~\cite{Romero_2018, Claudino_2020}. 
However, a separate study on the optimization strategy in {the} presence of noise is required. 
In this work, we use gradient-based Sequential Least SQuares Programming (SLSQP) optimizer~\cite{Kraft_1988}. We set {a} maximum number of iterations to 200 and stricted ftol to $10^{-10}$ in SLSQP optimizer for the considered molecules except for CO$_2$ to avoid optimization difficulties.

In ADAPT-VQE, we calculate gradients in the VQE procedure and gradients with respect to the operators in the same way. Such implementation is appropriate for real devices and lassical simulators, making it a general approach.
For the qubit operators, we calculate gradients directly using the parameter-shift rule~\cite{Schuld_2019}, which requires two circuit evaluations per one-parameter derivative. 
It is possible because qubit operators are represented by individual Pauli strings. 

Fermionic operators do not allow direct application of the parameter-shift rule. 
Analytic gradients in the fermionic ADAPT-VQE and VQE-UCCSD can be computed with the recently proposed fermionic-shift gradients~\cite{Kottmann} with 4 circuit evaluations per gradient or 2 evaluations introducing approximations. 
{However, each circuit evaluation takes significant time because of the large gate count in fermionic methods and large molecules used for analysis (up to 19 qubits).}
Since we perform statevector simulations, we use numerical gradients for fermionic methods, as they are more efficient in terms of circuit evaluations: 
only~$k + 1$ circuit evaluations are performed, where $k$ is the number of parameters in a quantum circuit.

In the case of ideal (noiseless) simulations, numerical and analytic gradients exactly match each other, while numerical ones are less robust in the presence of noise.
Analytic gradients, which we use for qubit methods, are important for future studies of qubit ADAPT-VQE performance in noisy conditions.

\subsection{Convergence criteria}\label{subsec:convergence_criteria}

In our simulation, we run ADAPT-VQE procedure until the energy convergence. 
However, it is preferable to use some selected convergence criteria in practical simulation. 
In the original work~\cite{Grimsley}, the authors use a convergence threshold on the norm of the gradient vector:
\begin{equation}
    \varepsilon_m = 10^{-m}.
\end{equation}

Other options are also possible. 
Specifically, the threshold can be applied to the maximum gradient element instead of the gradient vector norm. 
According to our calculations, setting the threshold for {the} maximum derivative to $10^{-2}$ provides ADAPT-VQE energies close to the VQE-UCCSD results but with improved gate count. 

\section*{Acknowledgements}
The research is supported by the RSF (Grant No. 19-71-10092; part related to simulating LiH and H$_2$O molecules using various VQE-based approaches) as well as Nissan Research Division, Nissan Motor Co., Ltd. and Nissan Research Center-Russia. We are grateful to Dr. Atsushi Oma, Nissan Research Division, Nissan Motor Co., Ltd., for useful discussions.
The development of the computational scheme for complete pools was supported by the Leading Research Center program (Agreement No. 014/20)
We also acknowledge the use of Qiskit Library~\cite{Qiskit} and QASM simulator.

\section*{Author contributions}
The authors jointly developed the problem statement 
and analyzed existing state-of-the-art techniques. 
M.S. implemented the considered methods and performed the simulation of molecules.
The authors jointly analyzed the results and wrote the manuscript.

\section*{Competing interests}
Owing to the employments and consulting activities of authors, they have financial interests in the commercial applications of quantum computing.

\section*{Code availability}
The code that is deemed central to the conclusions is available from the corresponding author upon reasonable request.

\bibliographystyle{naturemag}
\bibliography{CO2bib}

\begin{thebibliography}{10}
\expandafter\ifx\csname url\endcsname\relax
  \def\url#1{\texttt{#1}}\fi
\expandafter\ifx\csname urlprefix\endcsname\relax\def\urlprefix{URL }\fi
\providecommand{\bibinfo}[2]{#2}
\providecommand{\eprint}[2][]{\url{#2}}

\bibitem{Lloyd}
\bibinfo{author}{Lloyd, S.}
\newblock \bibinfo{title}{Universal quantum simulators}.
\newblock \emph{\bibinfo{journal}{Science}} \textbf{\bibinfo{volume}{273}},
  \bibinfo{pages}{1073--1078} (\bibinfo{year}{1996}).

\bibitem{McArdle}
\bibinfo{author}{McArdle, S.}, \bibinfo{author}{Endo, S.},
  \bibinfo{author}{Aspuru-Guzik, A.}, \bibinfo{author}{Benjamin, S.~C.} \&
  \bibinfo{author}{Yuan, X.}
\newblock \bibinfo{title}{Quantum computational chemistry}.
\newblock \emph{\bibinfo{journal}{Rev. Mod. Phys.}}
  \textbf{\bibinfo{volume}{92}}, \bibinfo{pages}{015003}
  (\bibinfo{year}{2020}).
\newblock
  \urlprefix\url{https://link.aps.org/doi/10.1103/RevModPhys.92.015003}.

\bibitem{Bauer_2020}
\bibinfo{author}{Bauer, B.}, \bibinfo{author}{Bravyi, S.},
  \bibinfo{author}{Motta, M.} \& \bibinfo{author}{Chan, G. K.-L.}
\newblock \bibinfo{title}{Quantum algorithms for quantum chemistry and quantum
  materials science}.
\newblock \emph{\bibinfo{journal}{Chemical Reviews}}
  \textbf{\bibinfo{volume}{120}}, \bibinfo{pages}{12685–12717}
  (\bibinfo{year}{2020}).
\newblock \urlprefix\url{http://dx.doi.org/10.1021/acs.chemrev.9b00829}.

\bibitem{elfving2020quantum}
\bibinfo{author}{Elfving, V.~E.} \emph{et~al.}
\newblock \bibinfo{title}{How will quantum computers provide an industrially
  relevant computational advantage in quantum chemistry?}
  (\bibinfo{year}{2020}).
\newblock \eprint{2009.12472}.

\bibitem{Fedorov2021}
\bibinfo{author}{Fedorov, A.~K.} \& \bibinfo{author}{Gelfand, M.~S.}
\newblock \bibinfo{title}{Towards practical applications in quantum
  computational biology}.
\newblock \emph{\bibinfo{journal}{Nature Computational Science}}
  \textbf{\bibinfo{volume}{1}}, \bibinfo{pages}{114--119}
  (\bibinfo{year}{2021}).
\newblock \urlprefix\url{https://doi.org/10.1038/s43588-021-00024-z}.

\bibitem{Cerezo_2021}
\bibinfo{author}{Cerezo, M. e.~a.}
\newblock \bibinfo{title}{Variational quantum algorithms}.
\newblock \emph{\bibinfo{journal}{Nature Reviews Physics}}
  \textbf{\bibinfo{volume}{38}} (\bibinfo{year}{2021}).
\newblock \urlprefix\url{https://doi.org/10.1038/s42254-021-00348-9}.

\bibitem{Peruzzo}
\bibinfo{author}{Peruzzo, A.} \emph{et~al.}
\newblock \bibinfo{title}{A variational eigenvalue solver on a photonic quantum
  processor}.
\newblock \emph{\bibinfo{journal}{Nature Communications}}
  \textbf{\bibinfo{volume}{5}}, \bibinfo{pages}{4213} (\bibinfo{year}{2014}).
\newblock \urlprefix\url{https://doi.org/10.1038/ncomms5213}.

\bibitem{fedorov2021vqe}
\bibinfo{author}{Fedorov, D.~A.}, \bibinfo{author}{Peng, B.},
  \bibinfo{author}{Govind, N.} \& \bibinfo{author}{Alexeev, Y.}
\newblock \bibinfo{title}{Vqe method: A short survey and recent developments}
  (\bibinfo{year}{2021}).
\newblock \eprint{2103.08505}.

\bibitem{Kandala2017}
\bibinfo{author}{Kandala, A.} \emph{et~al.}
\newblock \bibinfo{title}{Hardware-efficient variational quantum eigensolver
  for small molecules and quantum magnets}.
\newblock \emph{\bibinfo{journal}{Nature}} \textbf{\bibinfo{volume}{549}},
  \bibinfo{pages}{242--246} (\bibinfo{year}{2017}).

\bibitem{Barkoutsos}
\bibinfo{author}{Barkoutsos, P.~K.} \emph{et~al.}
\newblock \bibinfo{title}{Quantum algorithms for electronic structure
  calculations: Particle-hole hamiltonian and optimized wave-function
  expansions}.
\newblock \emph{\bibinfo{journal}{Phys. Rev. A}} \textbf{\bibinfo{volume}{98}},
  \bibinfo{pages}{022322} (\bibinfo{year}{2018}).
\newblock \urlprefix\url{https://link.aps.org/doi/10.1103/PhysRevA.98.022322}.

\bibitem{Gard}
\bibinfo{author}{Gard, B.} \emph{et~al.}
\newblock \bibinfo{title}{Efficient symmetry-preserving state preparation
  circuits for the variational quantum eigensolver algorithm}.
\newblock \emph{\bibinfo{journal}{npj Quantum Information}}
  \textbf{\bibinfo{volume}{6}} (\bibinfo{year}{2020}).

\bibitem{McClean}
\bibinfo{author}{Mcclean, J.}, \bibinfo{author}{Boixo, S.},
  \bibinfo{author}{Smelyanskiy, V.}, \bibinfo{author}{Babbush, R.} \&
  \bibinfo{author}{Neven, H.}
\newblock \bibinfo{title}{Barren plateaus in quantum neural network training
  landscapes}.
\newblock \emph{\bibinfo{journal}{Nature Communications}}
  \textbf{\bibinfo{volume}{9}} (\bibinfo{year}{2018}).

\bibitem{Romero_2018}
\bibinfo{author}{Romero, J.} \emph{et~al.}
\newblock \bibinfo{title}{Strategies for quantum computing molecular energies
  using the unitary coupled cluster ansatz}.
\newblock \emph{\bibinfo{journal}{Quantum Science and Technology}}
  \textbf{\bibinfo{volume}{4}}, \bibinfo{pages}{014008} (\bibinfo{year}{2018}).
\newblock \urlprefix\url{https://doi.org/10.1088/2058-9565/aad3e4}.

\bibitem{Kuhn}
\bibinfo{author}{Kühn, M.}, \bibinfo{author}{Zanker, S.},
  \bibinfo{author}{Deglmann, P.}, \bibinfo{author}{Marthaler, M.} \&
  \bibinfo{author}{Weiss, H.}
\newblock \bibinfo{title}{Accuracy and resource estimations for quantum
  chemistry on a near-term quantum computer}.
\newblock \emph{\bibinfo{journal}{Journal of Chemical Theory and Computation}}
  \textbf{\bibinfo{volume}{15}} (\bibinfo{year}{2019}).

\bibitem{armaos2019computational}
\bibinfo{author}{Armaos, V.}, \bibinfo{author}{Badounas, D.~A.} \&
  \bibinfo{author}{Deligiannis, P.}
\newblock \bibinfo{title}{Computational chemistry on quantum computers: Ground
  state estimation} (\bibinfo{year}{2019}).
\newblock \eprint{1907.00362}.

\bibitem{Rice}
\bibinfo{author}{Rice, J.} \emph{et~al.}
\newblock \bibinfo{title}{Quantum computation of dominant products in
  lithium–sulfur batteries}.
\newblock \emph{\bibinfo{journal}{The Journal of Chemical Physics}}
  \textbf{\bibinfo{volume}{154}}, \bibinfo{pages}{134115}
  (\bibinfo{year}{2021}).

\bibitem{OMalley_2016}
\bibinfo{author}{O'Malley, P. J.~J.} \emph{et~al.}
\newblock \bibinfo{title}{Scalable quantum simulation of molecular energies}.
\newblock \emph{\bibinfo{journal}{Phys. Rev. X}} \textbf{\bibinfo{volume}{6}},
  \bibinfo{pages}{031007} (\bibinfo{year}{2016}).
\newblock \urlprefix\url{https://link.aps.org/doi/10.1103/PhysRevX.6.031007}.

\bibitem{Hempel}
\bibinfo{author}{Hempel, C.} \emph{et~al.}
\newblock \bibinfo{title}{Quantum chemistry calculations on a trapped-ion
  quantum simulator}.
\newblock \emph{\bibinfo{journal}{Phys. Rev. X}} \textbf{\bibinfo{volume}{8}},
  \bibinfo{pages}{031022} (\bibinfo{year}{2018}).
\newblock \urlprefix\url{https://link.aps.org/doi/10.1103/PhysRevX.8.031022}.

\bibitem{McCaskey_2019}
\bibinfo{author}{McCaskey, A.} \emph{et~al.}
\newblock \bibinfo{title}{Quantum chemistry as a benchmark for near-term
  quantum computers}.
\newblock \emph{\bibinfo{journal}{npj Quantum Information}}
  \textbf{\bibinfo{volume}{5}} (\bibinfo{year}{2019}).

\bibitem{Grimsley_Trot}
\bibinfo{author}{Grimsley, H.~R.}, \bibinfo{author}{Claudino, D.},
  \bibinfo{author}{Economou, S.~E.}, \bibinfo{author}{Barnes, E.} \&
  \bibinfo{author}{Mayhall, N.~J.}
\newblock \bibinfo{title}{Is the trotterized uccsd ansatz chemically
  well-defined?}
\newblock \emph{\bibinfo{journal}{Journal of Chemical Theory and Computation}}
  \textbf{\bibinfo{volume}{16}}, \bibinfo{pages}{1--6} (\bibinfo{year}{2020}).
\newblock \urlprefix\url{https://doi.org/10.1021/acs.jctc.9b01083}.

\bibitem{Cooper_2010}
\bibinfo{author}{Cooper, B.} \& \bibinfo{author}{Knowles, P.}
\newblock \bibinfo{title}{Benchmark studies of variational, unitary and
  extended coupled cluster methods}.
\newblock \emph{\bibinfo{journal}{The Journal of chemical physics}}
  \textbf{\bibinfo{volume}{133}}, \bibinfo{pages}{234102}
  (\bibinfo{year}{2010}).

\bibitem{Lee}
\bibinfo{author}{Lee, J.}, \bibinfo{author}{Huggins, W.},
  \bibinfo{author}{Head-Gordon, M.} \& \bibinfo{author}{Whaley, K.}
\newblock \bibinfo{title}{Generalized unitary coupled cluster wavefunctions for
  quantum computation}.
\newblock \emph{\bibinfo{journal}{Journal of Chemical Theory and Computation}}
  \textbf{\bibinfo{volume}{15}} (\bibinfo{year}{2018}).

\bibitem{Helgaker_1997}
\bibinfo{author}{Helgaker, T.}, \bibinfo{author}{Klopper, W.} \&
  \bibinfo{author}{Koch, H.}
\newblock \bibinfo{title}{Basis-set convergence of correlated calculations on
  water}.
\newblock \emph{\bibinfo{journal}{The Journal of Chemical Physics}}
  \textbf{\bibinfo{volume}{106}}, \bibinfo{pages}{9639--9646}
  (\bibinfo{year}{1997}).

\bibitem{Varandas_2018}
\bibinfo{author}{Varandas, A.}
\newblock \bibinfo{title}{Straightening the hierarchical staircase for basis
  set extrapolations: A low-cost approach to high-accuracy computational
  chemistry}.
\newblock \emph{\bibinfo{journal}{Annual Review of Physical Chemistry}}
  \textbf{\bibinfo{volume}{69}} (\bibinfo{year}{2018}).

\bibitem{Musia2001CoupledCS}
\bibinfo{author}{Musiał, M.}, \bibinfo{author}{Kucharski, S.} \&
  \bibinfo{author}{Bartlett, R.}
\newblock \bibinfo{title}{Coupled cluster study of the triple bond}.
\newblock \emph{\bibinfo{journal}{Journal of Molecular Structure-theochem}}
  \textbf{\bibinfo{volume}{547}}, \bibinfo{pages}{269--278}
  (\bibinfo{year}{2001}).

\bibitem{Helgaker2001}
\bibinfo{author}{Helgaker, T.} \emph{et~al.}
\newblock \emph{\bibinfo{title}{Highly Accurate Ab Initio Computation of
  Thermochemical Data}}, \bibinfo{pages}{1--30} (\bibinfo{publisher}{Springer
  Netherlands}, \bibinfo{address}{Dordrecht}, \bibinfo{year}{2001}).
\newblock \urlprefix\url{https://doi.org/10.1007/0-306-47632-0_1}.

\bibitem{Grimsley}
\bibinfo{author}{Grimsley, H.~R.}, \bibinfo{author}{Economou, S.~E.},
  \bibinfo{author}{Barnes, E.} \& \bibinfo{author}{Mayhall, N.~J.}
\newblock \bibinfo{title}{An adaptive variational algorithm for exact molecular
  simulations on a quantum computer}.
\newblock \emph{\bibinfo{journal}{Nature Communications}}
  \textbf{\bibinfo{volume}{10}}, \bibinfo{pages}{3007} (\bibinfo{year}{2019}).
\newblock \urlprefix\url{https://doi.org/10.1038/s41467-019-10988-2}.

\bibitem{Tang_2021}
\bibinfo{author}{Tang, H.} \emph{et~al.}
\newblock \bibinfo{title}{Qubit-adapt-vqe: An adaptive algorithm for
  constructing hardware-efficient ansätze on a quantum processor}.
\newblock \emph{\bibinfo{journal}{PRX Quantum}} \textbf{\bibinfo{volume}{2}}
  (\bibinfo{year}{2021}).

\bibitem{shkolnikov2021avoiding}
\bibinfo{author}{Shkolnikov, V.~O.}, \bibinfo{author}{Mayhall, N.~J.},
  \bibinfo{author}{Economou, S.~E.} \& \bibinfo{author}{Barnes, E.}
\newblock \bibinfo{title}{Avoiding symmetry roadblocks and minimizing the
  measurement overhead of adaptive variational quantum eigensolvers}
  (\bibinfo{year}{2021}).
\newblock \eprint{2109.05340}.

\bibitem{Liu2021AnEA}
\bibinfo{author}{Liu, J.}, \bibinfo{author}{Li, Z.} \& \bibinfo{author}{Yang,
  J.}
\newblock \bibinfo{title}{An efficient adaptive variational quantum solver of
  the schr{\"o}dinger equation based on reduced density matrices.}
\newblock \emph{\bibinfo{journal}{The Journal of chemical physics}}
  \textbf{\bibinfo{volume}{154 24}}, \bibinfo{pages}{244112}
  (\bibinfo{year}{2021}).

\bibitem{Evangelista_2019}
\bibinfo{author}{Evangelista, F.}, \bibinfo{author}{Chan, G.} \&
  \bibinfo{author}{Scuseria, G.}
\newblock \bibinfo{title}{Exact parameterization of fermionic wave functions
  via unitary coupled cluster theory}.
\newblock \emph{\bibinfo{journal}{The journal of chemical physics}}
  (\bibinfo{year}{2019}).

\bibitem{Setia_2020}
\bibinfo{author}{Setia, K.} \emph{et~al.}
\newblock \bibinfo{title}{Reducing qubit requirements for quantum simulations
  using molecular point group symmetries}.
\newblock \emph{\bibinfo{journal}{Journal of Chemical Theory and Computation}}
  \textbf{\bibinfo{volume}{16}}, \bibinfo{pages}{6091--6097}
  (\bibinfo{year}{2020}).
\newblock \urlprefix\url{https://doi.org/10.1021/acs.jctc.0c00113}.

\bibitem{Soubaihi_CO_oxid}
\bibinfo{author}{Al~Soubaihi, R.~M.}, \bibinfo{author}{Saoud, K.~M.} \&
  \bibinfo{author}{Dutta, J.}
\newblock \bibinfo{title}{Critical review of low-temperature co oxidation and
  hysteresis phenomenon on heterogeneous catalysts}.
\newblock \emph{\bibinfo{journal}{Catalysts}} \textbf{\bibinfo{volume}{8}}
  (\bibinfo{year}{2018}).
\newblock \urlprefix\url{https://www.mdpi.com/2073-4344/8/12/660}.

\bibitem{Dey}
\bibinfo{author}{Dey, S.} \& \bibinfo{author}{Dhal, G.}
\newblock \bibinfo{title}{Catalytic conversion of carbon monoxide into carbon
  dioxide over spinel catalysts: An overview}.
\newblock \emph{\bibinfo{journal}{Materials Science for Energy Technologies}}
  \textbf{\bibinfo{volume}{2}} (\bibinfo{year}{2019}).

\bibitem{Ryabinkin_2018}
\bibinfo{author}{Ryabinkin, I.}, \bibinfo{author}{Yen, T.-C.},
  \bibinfo{author}{Genin, S.} \& \bibinfo{author}{Izmaylov, A.}
\newblock \bibinfo{title}{Qubit coupled cluster method: A systematic approach
  to quantum chemistry on a quantum computer}.
\newblock \emph{\bibinfo{journal}{Journal of Chemical Theory and Computation}}
  \textbf{\bibinfo{volume}{14}} (\bibinfo{year}{2018}).

\bibitem{Ryabinkin_2020}
\bibinfo{author}{Ryabinkin, I.~G.}, \bibinfo{author}{Lang, R.~A.},
  \bibinfo{author}{Genin, S.~N.} \& \bibinfo{author}{Izmaylov, A.~F.}
\newblock \bibinfo{title}{Iterative qubit coupled cluster approach with
  efficient screening of generators.}
\newblock \emph{\bibinfo{journal}{Journal of chemical theory and computation}}
  (\bibinfo{year}{2020}).

\bibitem{Qiskit}
\bibinfo{author}{Md~Sajid~Anis, H. A. e.~a.}
\newblock \bibinfo{title}{Qiskit: An open-source framework for quantum
  computing} (\bibinfo{year}{2021}).

\bibitem{Soini_2015}
\bibinfo{author}{Soini, T.}, \bibinfo{author}{Genest, A.} \&
  \bibinfo{author}{Rösch, N.}
\newblock \bibinfo{title}{Assessment of hybrid density functionals for the
  adsorption of carbon monoxide on platinum model clusters}.
\newblock \emph{\bibinfo{journal}{The journal of physical chemistry. A}}
  \textbf{\bibinfo{volume}{119}} (\bibinfo{year}{2015}).

\bibitem{Feibelman_2001}
\bibinfo{author}{Feibelman, P. J. e.~a.}
\newblock \bibinfo{title}{The co/pt(111) puzzle}.
\newblock \emph{\bibinfo{journal}{The Journal of Physical Chemistry B}}
  \textbf{\bibinfo{volume}{105}}, \bibinfo{pages}{4018--4025}
  (\bibinfo{year}{2001}).
\newblock \urlprefix\url{https://doi.org/10.1021/jp002302t}.
\newblock \eprint{https://doi.org/10.1021/jp002302t}.

\bibitem{Douglas_Gallardo_2019}
\bibinfo{author}{Douglas-Gallardo, O.~A.}, \bibinfo{author}{Saez, D.~A.},
  \bibinfo{author}{Vogt-Geisse, S.} \& \bibinfo{author}{Vöhringer-Martinez,
  E.}
\newblock \bibinfo{title}{Electronic structure benchmark calculations of
  inorganic and biochemical carboxylation reactions}.
\newblock \emph{\bibinfo{journal}{Journal of Computational Chemistry}}
  \textbf{\bibinfo{volume}{40}}, \bibinfo{pages}{1401--1413}
  (\bibinfo{year}{2019}).
\newblock
  \urlprefix\url{https://onlinelibrary.wiley.com/doi/abs/10.1002/jcc.25795}.
\newblock \eprint{https://onlinelibrary.wiley.com/doi/pdf/10.1002/jcc.25795}.

\bibitem{JordanberDP}
\bibinfo{author}{Jordan, P.} \& \bibinfo{author}{Wigner, E.}
\newblock \bibinfo{title}{{\"U}ber das paulische {\"a}quivalenzverbot}.
\newblock \emph{\bibinfo{journal}{Zeitschrift f{\"u}r Physik}}
  \textbf{\bibinfo{volume}{47}}, \bibinfo{pages}{631--651}.

\bibitem{Yordanov_fermionic}
\bibinfo{author}{Yordanov, Y.~S.}, \bibinfo{author}{Arvidsson-Shukur, D. R.~M.}
  \& \bibinfo{author}{Barnes, C. H.~W.}
\newblock \bibinfo{title}{Efficient quantum circuits for quantum computational
  chemistry}.
\newblock \emph{\bibinfo{journal}{Phys. Rev. A}}
  \textbf{\bibinfo{volume}{102}}, \bibinfo{pages}{062612}
  (\bibinfo{year}{2020}).
\newblock \urlprefix\url{https://link.aps.org/doi/10.1103/PhysRevA.102.062612}.

\bibitem{Yordanoc_iqeb}
\bibinfo{author}{Yordanov, Y.}, \bibinfo{author}{Armaos, V.},
  \bibinfo{author}{Barnes, C.} \& \bibinfo{author}{Shukur, D.}
\newblock \bibinfo{title}{Iterative qubit-excitation based variational quantum
  eigensolver}  (\bibinfo{year}{2020}).

\bibitem{Sun_2020}
\bibinfo{author}{Sun, Q.} \emph{et~al.}
\newblock \bibinfo{title}{Recent developments in the pyscf program package}.
\newblock \emph{\bibinfo{journal}{The Journal of Chemical Physics}}
  \textbf{\bibinfo{volume}{153}}, \bibinfo{pages}{024109}
  (\bibinfo{year}{2020}).
\newblock \urlprefix\url{https://doi.org/10.1063/5.0006074}.
\newblock \eprint{https://doi.org/10.1063/5.0006074}.

\bibitem{Claudino_2020}
\bibinfo{author}{Claudino, D.}, \bibinfo{author}{Wright, J.},
  \bibinfo{author}{McCaskey, A.} \& \bibinfo{author}{Humble, T.}
\newblock \bibinfo{title}{Benchmarking adaptive variational quantum
  eigensolvers}.
\newblock \emph{\bibinfo{journal}{Frontiers in Chemistry}}
  \textbf{\bibinfo{volume}{8}} (\bibinfo{year}{2020}).

\bibitem{Kraft_1988}
\bibinfo{author}{Kraft, D.}
\newblock \emph{\bibinfo{title}{A software package for sequential quadratic
  programming}}.
\newblock Deutsche Forschungs- und Versuchsanstalt f{\"u}r Luft- und Raumfahrt
  K{\"o}ln: Forschungsbericht (\bibinfo{publisher}{Wiss. Berichtswesen d.
  DFVLR}, \bibinfo{year}{1988}).
\newblock \urlprefix\url{https://books.google.ru/books?id=4rKaGwAACAAJ}.

\bibitem{Schuld_2019}
\bibinfo{author}{Schuld, M.}, \bibinfo{author}{Bergholm, V.},
  \bibinfo{author}{Gogolin, C.}, \bibinfo{author}{Izaac, J.} \&
  \bibinfo{author}{Killoran, N.}
\newblock \bibinfo{title}{Evaluating analytic gradients on quantum hardware}.
\newblock \emph{\bibinfo{journal}{Physical Review A}}
  \textbf{\bibinfo{volume}{99}} (\bibinfo{year}{2019}).
\newblock \urlprefix\url{http://dx.doi.org/10.1103/PhysRevA.99.032331}.

\bibitem{Kottmann}
\bibinfo{author}{Kottmann, J.~S.}, \bibinfo{author}{Anand, A.} \&
  \bibinfo{author}{Aspuru-Guzik, A.}
\newblock \bibinfo{title}{A feasible approach for automatically differentiable
  unitary coupled-cluster on quantum computers}.
\newblock \emph{\bibinfo{journal}{Chem. Sci.}} \textbf{\bibinfo{volume}{12}},
  \bibinfo{pages}{3497--3508} (\bibinfo{year}{2021}).
\newblock \urlprefix\url{http://dx.doi.org/10.1039/D0SC06627C}.

\end{thebibliography}

\newpage\newpage

\begin{widetext} 
\renewcommand{\thesection}{S\arabic{section}} 
\renewcommand{\theequation}{S\arabic{equation}}
\renewcommand{\thefigure}{S\arabic{figure}}
\renewcommand{\thetable}{S\arabic{table}}
\setcounter{equation}{0}
\setcounter{figure}{0}
\setcounter{table}{0}
\newpage\newpage
\hrulefill
\begin{center}
\textbf{SUPPLEMENTARY INFORMATION}
\end{center}

\begin{figure*}[htbp]
\centering
    \includegraphics[width=0.9\textwidth]{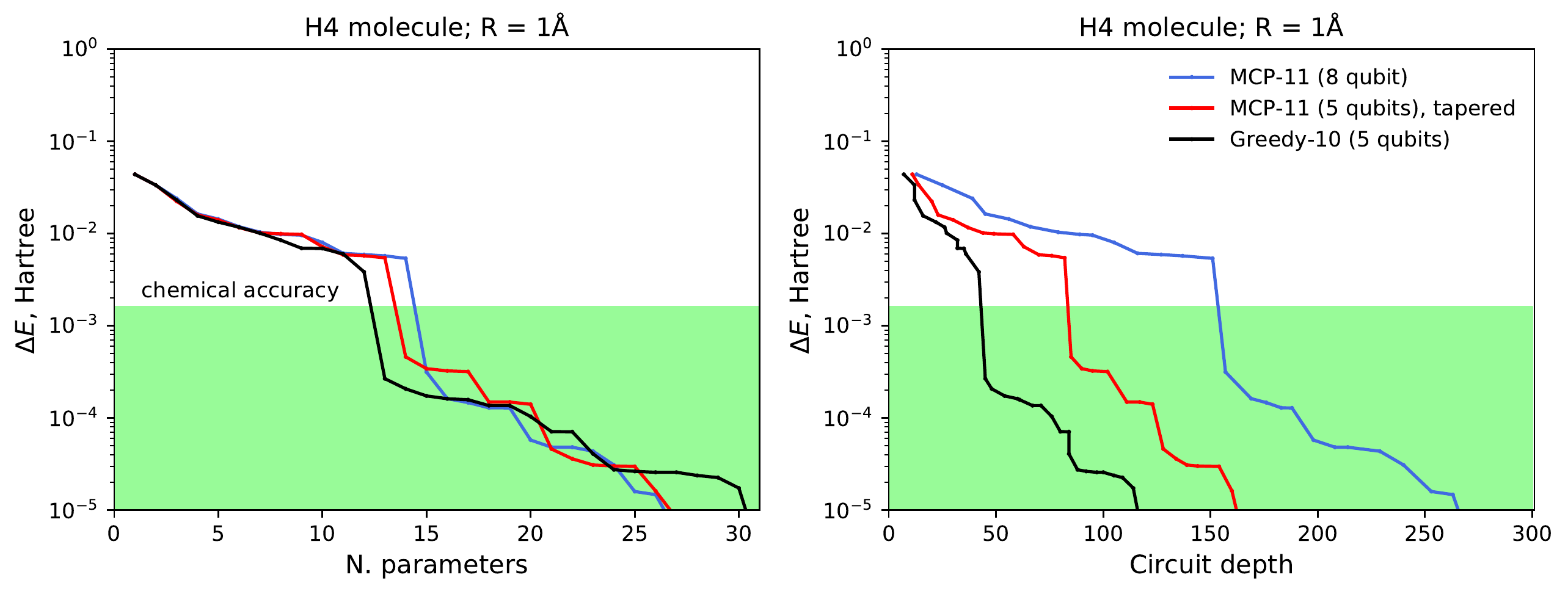}
    \centering
    \caption{\textbf{Qubit ADAPT-VQE convergence of energy with respect to the number of parameters and the circuit depth for H$_4$ molecule at $R = 1.0${\AA}.} ``MCP-11'' is the pool acting in the original qubit space of 8 qubits suggested in previous work~\cite{shkolnikov2021avoiding}. ``MCP-11, tapered'' is built from ``MCP-11'' by applying tapering procedure. ``Greedy-10'' pool is constructed using the computational scheme proposed in this work and acts in a 5-qubit space.}
    \label{fig:h4_adapt-vqe-conv_r1}
\end{figure*}

\begin{figure*}[htbp]
\centering
    \includegraphics[width=0.9\textwidth]{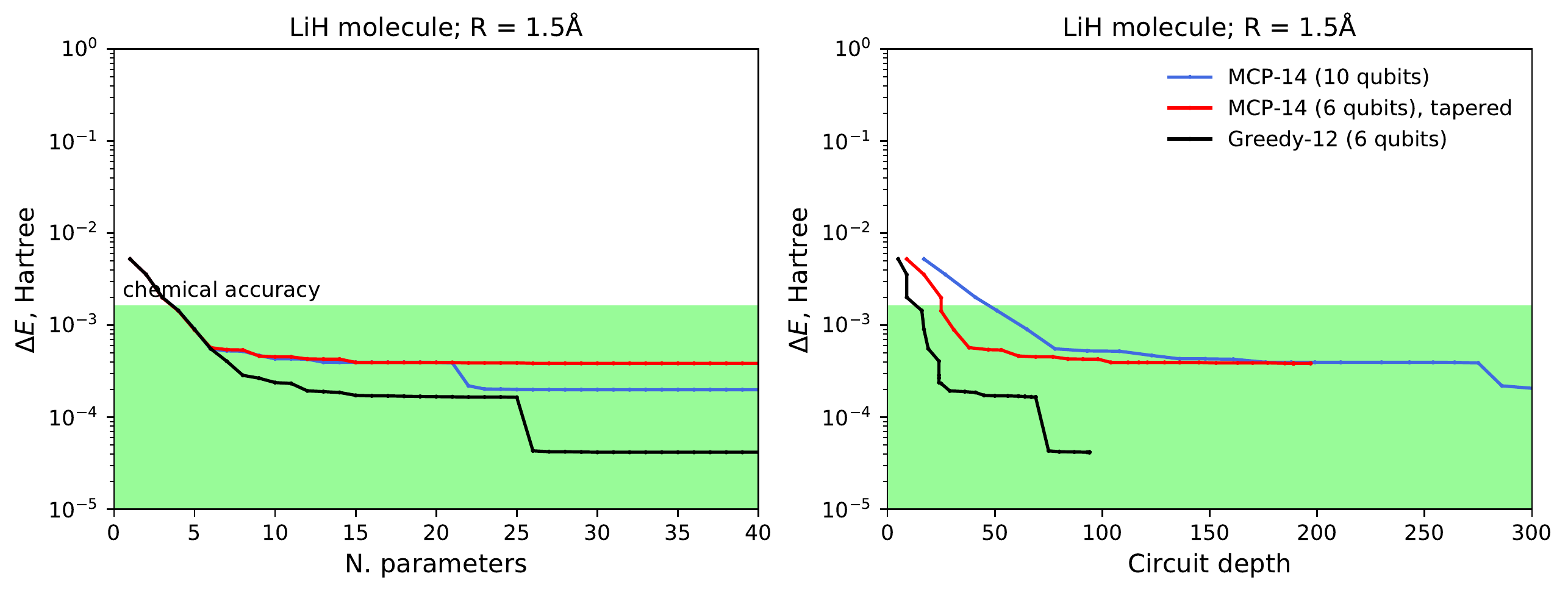}
    \centering
    \caption{\textbf{Qubit ADAPT-VQE convergence of energy with respect to the number of parameters and the circuit depth for LiH molecule at $R = 1.5${\AA}.} ``MCP-14'' is the pool acting in the original qubit space of 10 qubits suggested in previous work~\cite{shkolnikov2021avoiding}. ``MCP-14, tapered'' is built from ``MCP-14'' by applying tapering procedure. ``Greedy-12'' pool is constructed using the computational scheme proposed in this work and acts in a 6-qubit space.}
    \label{fig:lih-convergence_r1}
\end{figure*}

\begin{figure}[htbp]
\centering
    \includegraphics[width=0.5\linewidth]{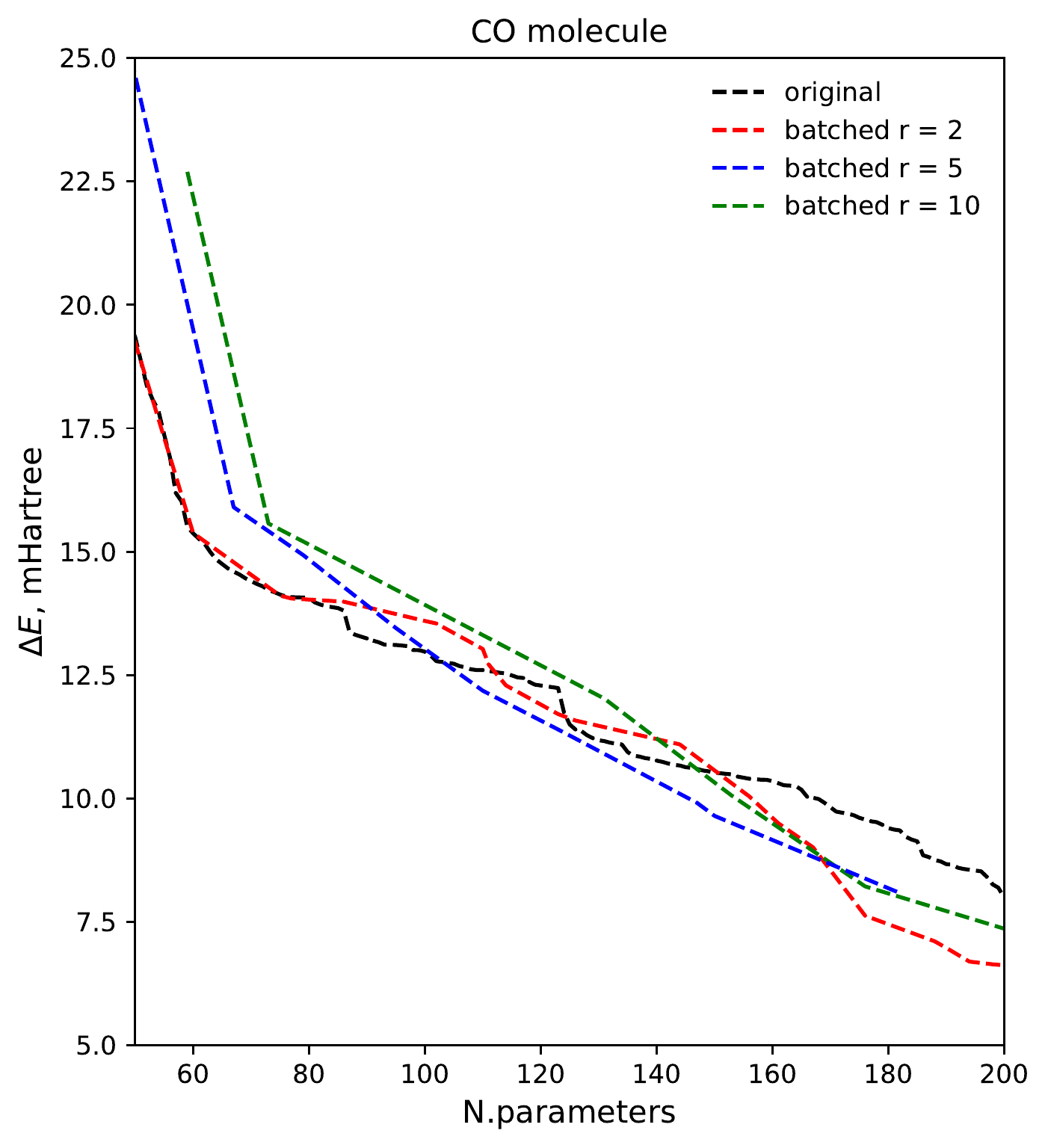}
    \centering
    \caption{\textbf{Energy convergence for CO molecule for batched and original qubit ADAPT-VQE}. Batched implementation was tested with gradient ratio $r = 2, 5, 10$. We selected $r = 2$ for impelementation of batched ADAPT-VQE for all the considered molecules.}
    \label{fig:CO_gradratio}
\end{figure}

\begin{figure}[htbp]
\centering
    \includegraphics[width=0.6\linewidth]{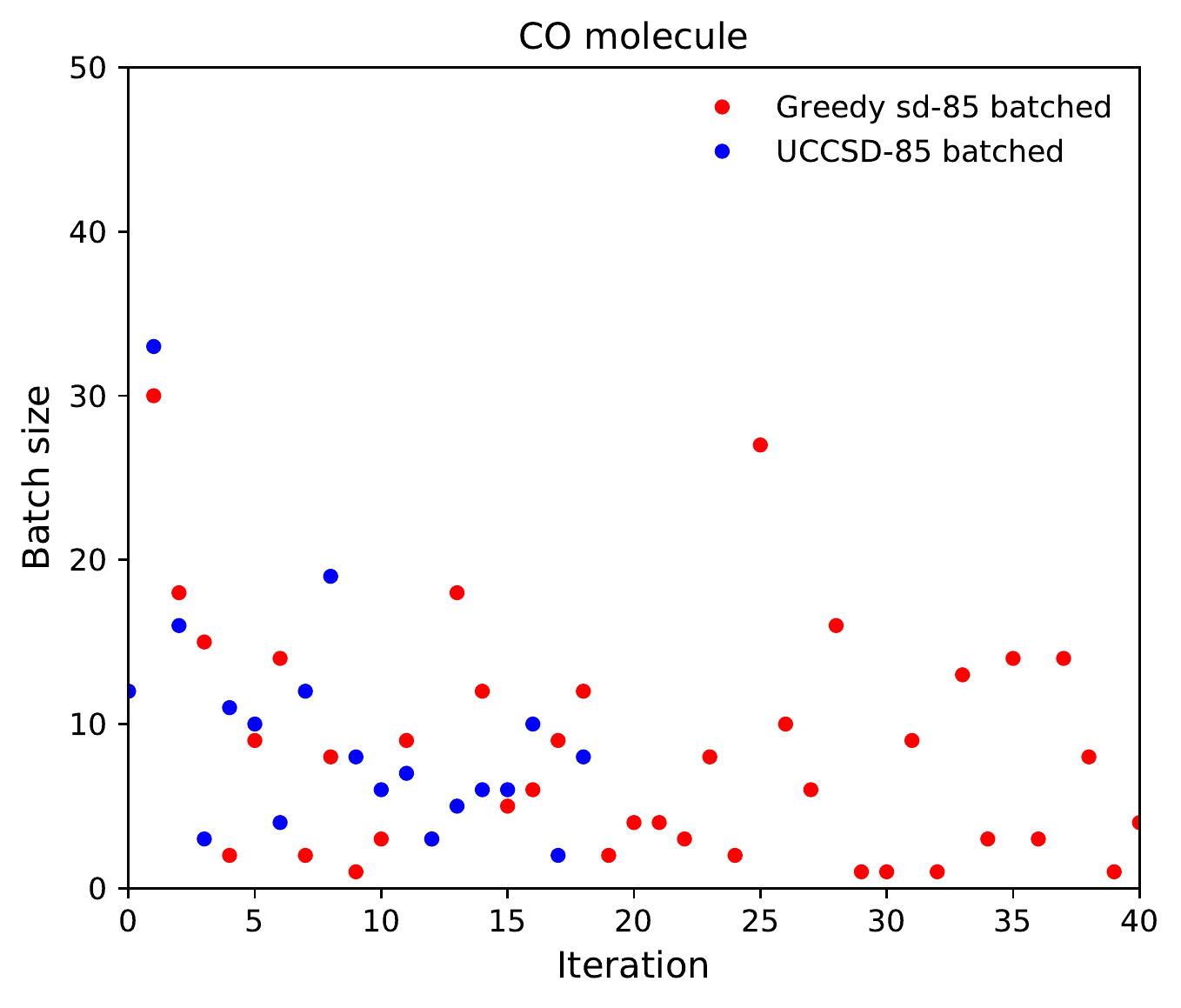}
    \centering
    \caption{\textbf{Batch size at different iterations of batched ADAPT-VQE for CO molecules.}}
    \label{fig:CO_gradratio}
\end{figure}

\end{widetext}

\end{document}